\documentclass[pre,aps,twocolumn,,amsmath,amssymb,showpacs,showkeys,floatfix]{revtex4-1}
\usepackage{amsfonts}
\usepackage{epsfig}
\usepackage{dcolumn}
\usepackage{amsthm}
\usepackage{color}
\usepackage{float}
\usepackage{latexsym}
\usepackage{amscd}
\usepackage{hyperref}
\usepackage[mathscr]{eucal}

\usepackage{bm}      
\usepackage{ifpdf}
\usepackage{bbm}

\begin{document}

\newcommand{\newc}{\newcommand}
\newc{\beq}{\begin{equation}}
\newc{\eeq}{\end{equation}}
\newc{\kt}{\rangle}
\newc{\br}{\langle}
\newc{\beqa}{\begin{eqnarray}}
\newc{\eeqa}{\end{eqnarray}}
\newc{\pr}{\prime}
\newc{\longra}{\longrightarrow}
\newc{\ot}{\otimes}
\newc{\rarrow}{\rightarrow}
\newc{\h}{\hat}
\newc{\bom}{\boldmath}
\newc{\btd}{\bigtriangledown}
\newc{\al}{\alpha}
\newc{\be}{\beta}
\newc{\ld}{\lambda}
\newc{\sg}{\sigma}
\newc{\p}{\psi}
\newc{\eps}{\epsilon}
\newc{\om}{\omega}
\newc{\mb}{\mbox}
\newc{\tm}{\times}
\newc{\hu}{\hat{u}}
\newc{\hv}{\hat{v}}
\newc{\hk}{\hat{K}}
\newc{\ra}{\rightarrow}
\newc{\non}{\nonumber}
\newc{\ul}{\underline}
\newc{\hs}{\hspace}
\newc{\longla}{\longleftarrow}
\newc{\ts}{\textstyle}
\newc{\f}{\frac}
\newc{\df}{\dfrac}
\newc{\ovl}{\overline}
\newc{\bc}{\begin{center}}
\newc{\ec}{\end{center}}
\newc{\dg}{\dagger}
\newc{\prh}{\mbox{PR}_H}
\newc{\prq}{\mbox{PR}_q}
\newc{\tr}{\mbox{tr}}
\newc{\pd}{\partial}
\newc{\qv}{\vec{q}}
\newc{\pv}{\vec{p}}
\newc{\dqv}{\delta\vec{q}}
\newc{\dpv}{\delta\vec{p}}
\newc{\mbq}{\mathbf{q}}
\newc{\mbqp}{\mathbf{q'}}
\newc{\mbpp}{\mathbf{p'}}
\newc{\mbp}{\mathbf{p}}
\newc{\mbn}{\mathbf{\nabla}}
\newc{\dmbq}{\delta \mbq}
\newc{\dmbp}{\delta \mbp}
\newc{\T}{\mathsf{T}}
\newc{\J}{\mathsf{J}}
\newc{\sfL}{\mathsf{L}}
\newc{\C}{\mathsf{C}}
\newc{\B}{\mathsf{M}}
\newc{\V}{\mathsf{V}}

\title{Large deviations induced entanglement transitions}
\author{Udaysinh T. Bhosale}
\email{udaybhosale0786@gmail.com}
\affiliation{Indian Institute of Science Education and Research, Dr. Homi Bhabha Road, Pune 411 008, India.}

\date{\today}

\begin{abstract}
 
The probability of large deviations of the smallest Schmidt eigenvalue for random pure states of bipartite systems, 
denoted as $A$ and $B$, is computed analytically using a Coulomb gas method. It is shown that this probability, for 
large $N$, goes as $\exp[-\beta N^2\Phi(\zeta)]$, where the parameter $\beta$ is the Dyson index of the ensemble,
$\zeta$ is the large deviation parameter while the rate function $\Phi(\zeta)$ is calculated exactly.
Corresponding equilibrium Coulomb charge density is derived for its large deviations.
Effects of the large deviations of the extreme (largest and smallest) Schmidt eigenvalues on the 
bipartite entanglement are studied using the von Neumann entropy. Effect of these deviations are also studied on the 
entanglement between subsystems $1$ and $2$, obtained by further partitioning the subsystem $A$,
using the properties of the density matrix's partial transpose $\rho_{12}^\Gamma$. The density of states of 
$\rho_{12}^\Gamma$ is found to be close to the Wigner's semicircle law with these large deviations. The entanglement 
properties are captured very well by a simple 
random matrix model for the partial transpose. The model predicts the entanglement transition across a critical
large deviation parameter $\zeta$. Log negativity is used to quantify the entanglement between subsystems $1$ and $2$. 
Analytical formulas for it are derived using the simple model.
Numerical simulations are in excellent agreement with the analytical results.

\end{abstract}

\maketitle

\section{Introduction}

The large deviation is defined as the atypical behavior of a system from its average state. Its theory is an active 
field of research in probability and statistics \cite{Hollander}. This theory has found applications in the 
field of random matrices \cite{Vivo07,Satyalarge,Satyalarge1,Castillo2016,Castillo2010,Vivo07,SatyaIndex2009}, 
quantum entanglement 
\cite{Nadal10,Arul08,majumdar08,Nadal11,Uday12,Karol2017,Vivo2011}, economics \cite{Chavez2015}, geophysics, 
hydrology \cite{Sergio2006}, image processing \cite{wilksbook,Fukunaga} etc. This theory is tested in the
context of coupled lasers and found to agree very well with the experiment \cite{Fridman12}.
It has been successfully applied in the field of quantum information to study entanglement.
Entanglement is a central property of quantum mechanics which is not there in classical physics. In fact recently 
it shown that any theory which has a classical limit must have entanglement as an inevitable feature 
\cite{Jonathan2017}. It is studied extensively since it is a critical resource for the quantum computation and 
information tasks \cite{Horodeckirpm}, quantum teleportation \cite{bennett93}, dense coding \cite{Superdense}, etc. 
Entanglement has been studied in various experiments using optics, superconductivity, etc \cite{Horodeckirpm}. 
In this paper, we are interested in the applications of the large deviation theory to study the 
entanglement transitions.  
 
Let us start by considering a standard bipartite system $A\otimes B$ which is composed of two smaller subsystems 
$A$ and $B$ having Hilbert spaces $\mathcal{H}{_A}{^{(N)}}$ and $\mathcal{H}{_B}{^{(M)}}$ having dimensions $N$ 
and $M$ respectively. Whereas the full system is described by the product Hilbert space 
$\mathcal{H}{_{AB}^{(MN)}}= \mathcal{H}{_A}{^{(N)}} \otimes \mathcal{H}{_B}{^{(M)}}$. 
Here, the simple case of $N=M$ is studied in detail but the results can be extended to the 
$N\neq M$ case. Consider $|\psi\rangle=\sum_{i=1}^N \sum_{\alpha=1}^M c_{i,\alpha} |i\rangle\otimes|\alpha\rangle$ 
a normalized pure state of the full system $A$ and $B$, where $|i\rangle\otimes|\alpha\rangle$ is the orthonormal 
basis of $\mathcal{H}{_{AB}}$. The density matrix is given as $\rho=|\psi\rangle\langle\psi|$ which satisfies 
Tr[$\rho$]=1 condition. The reduced density matrix of the subsystem $A$ is given by 
$\rho_A= \mbox{Tr}_B[\rho]=\sum_{\alpha=1}^{M} \langle \alpha| \rho|\alpha \rangle$. Similarly, the subsystem $B$ is 
described by $\rho_B= \mbox{Tr}_A[\rho]$.
Using the singular value decomposition of the matrix $c_{i,\alpha}$ one obtains the
Schmidt decomposition form:
\begin{equation}\label{Eq:SchmidtDecom}
 |\psi\rangle=\sum_{i=1}^{N}\sqrt{\lambda_i} | u_i^A\rangle \otimes | v_i^B \rangle
\end{equation}
where $|u_i^A\rangle$ and $|v_i^B \rangle$ are the eigenvectors of $\rho_A$ and $\rho_B$ respectively,
with the same eigenvalues $\lambda_i$. The $\lambda_i\in [0,1]$ for all $i=1$ to $N$ such that 
$\sum_{i=1}^N \lambda_i=1$.

Given the Schmidt eigenvalues $\lambda_i$ ($i=1\ldots N$), entanglement between $A$ and $B$, measured 
using von Neumann entropy, is given by 
\begin{equation}\label{Eq:vonNeumannEntropy}
S_{VN}=-\mbox{tr}(\rho_A\log\rho_A)=-\sum_{i=1}^{N}\lambda_i\ln(\lambda_i).
\end{equation} 
It is a good measure of entanglement for a bipartite pure state \cite{Bennett96,Zyczkowski06}. 
It takes value between $0$ which corresponds to separable state and $\ln(N)$ which corresponds to maximally
entangled state. Study of the two extreme eigenvalues, the largest 
$\lambda_{\mbox{max}}=\mbox{max}(\lambda_1,\lambda_2,...,\lambda_N)$ 
and the smallest $\lambda_{\mbox{min}}=\mbox{min}(\lambda_1,\lambda_2,...,\lambda_N)$, is important as they 
give useful information about the nature of entanglement between the subsystems $A$ and $B$
\cite{Demmel1988,Edelman1992,MarkoZ07,majumdar08,YangChen2010,Nadal10,Nadal11,Nadler2011}.
It can be seen easily that the conditions $\sum_{i=1}^N \lambda_i=1$ and $\lambda_i\in [0,1]$ for $i=1\ldots N$
imply $0\leq \lambda_{\mbox{min}}\leq 1/N$ and $1/N \leq \lambda_{\mbox{max}} \leq 1$.

To understand the importance of the extreme eigenvalues, let us first consider the following limiting situations 
of the largest eigenvalue. Suppose that $\lambda_{\mbox{max}}$
takes the maximum allowed value $1$. Then due to the normalization constraints $\sum_{i=1}^N \lambda_i=1$ and 
$\lambda_i\in [0,1]$ for all $i$, it follows that all the rest $(N-1)$ eigenvalues must be identically equal to $0$.
Thus, using Eq.~(\ref{Eq:SchmidtDecom}) for this case implies that the state $|\psi\rangle$ is fully 
{\it unentangled}. On the other hand, if $\lambda_{\mbox{max}}$ takes its lowest allowed value $1/N$
then the constraint $\sum_{i=1}^N \lambda_i=1$ implies that $\lambda_i=1/N$ for all $i$. In this case,
it can be shown that the state $|\psi\rangle$ is {\it maximally} entangled as it maximizes the von Neumann
entropy $S_{VN}=\ln(N)$.

Now, consider the limiting situations of the minimum eigenvalue. Suppose, 
$\lambda_{\mbox{min}}$ takes the maximum allowed value $1/N$. Then, the constraint $\sum_{i=1}^N \lambda_i=1$ 
implies that $\lambda_i=1/N$ for all $i$. Thus, the state $|\psi\rangle$ is {\it maximally} entangled.
When $\lambda_{\mbox{min}}$ takes the minimum allowed value $0$ then not much information on the entanglement in 
the state $|\psi\rangle$ is obtained. But, using the Schmidt decomposition one can see that the dimension of 
the effective Hilbert space of the subsystem $A$ is now reduced from $N$ to $N-1$. This also implies
that the maximum von Neumann entropy it can take is reduced from $\ln(N)$ to $\ln(N-1)$.

The pure state $|\psi\rangle$ is called random when it is sampled uniformly from the unique
Haar measure that is invariant under unitary transformations.
As a result, the eigenvalues $\lambda_i$'s also become random variables.
In that case, the distributions of the extreme eigenvalues of $\rho_A$ have been studied in detail for various 
cases of $N$ and $M$ \cite{Demmel1988,Edelman1992,MarkoZ07,majumdar08,YangChen2010,Nadal10,Nadal11,Nadler2011}.  
The distribution of the minimum eigenvalue for $\beta=1$, $2$ and finite $N=M$ was derived in 
\cite{Edelman1992,majumdar08} while $N \neq M$ case is addressed in \cite{YangChen2010}.
Here, $\beta$ is the Dyson index and it takes values $1$, $2$ and $4$ for real, complex and symplectic case 
respectively. Similarly, the maximum eigenvalue distribution for large $N=M$ and for all $\beta$s is given in 
\cite{Nadal10,Nadal11,Vivo2011}, which include the small and large deviation laws.
In fact, the distribution of all the Schmidt eigenvalues taken together for large $N$ and $M$ is known as the 
Marcenko-Pastur function
\cite{Nadal11,Marcenko} (see Eq.~(\ref{MPfunct1})).
Probability distribution of the Renyi entropies, a measure of entanglement, for a random pure state of a large 
bipartite quantum system has been derived analytically \cite{Nadal10,Nadal11,Borot2012}.

If the constraint of eigenvalues summing to one is removed and $c_{i,\alpha}$ are independent and identically 
distributed (i.i.d.) Gaussian random variables, real or complex, drawn from a Gaussian distribution, then $\rho_A$ belongs 
to the Wishart ensemble. These matrices have found applications in various fields like finance \cite{Bouchaud}, 
nuclear physics \cite{Fyodorov1997,Fyodorov1999}, quantum chromodynamics
\cite{Shuryak93,Verbaarschot94}, knowledge networks \cite{Maslov2001}, etc. For these ensembles it is shown that the 
probability distribution of the {\it typical and small} fluctuations of the extreme eigenvalues is given by 
Tracy-Widom distribution \cite{Tracy1,Tracy2,Satyalargeright}, while the {\it atypical and large} fluctuations 
obey a different distribution having limiting form of the Tracy-Widom in the limit of small fluctuations 
\cite{Vivo07,Castillo2010}. 

Turning our attention to $\rho_A$, whose eigenvalues are non-negative and sum to one, the large deviation function 
for the maximum eigenvalue and the corresponding equilibrium charge density
is derived in Ref.\cite{Nadal11} using the Coulomb gas method. To be specific, 
the probability distribution function $P(N\lambda_{\mbox{max}}= a)$ where $a>1$ is 
derived. It is also shown that its typical fluctuations around the average $4/N$ follow the Tracy-Widom Distribution.
In this paper the large deviation function for the minimum eigenvalue and the associated equilibrium charge density
is derived. Thus, a generalized Macenko-Pastur function is derived when there are large deviations in the minimum 
eigenvalue. 
For these derivations, the improved version of the coulomb gas technique from Ref.\cite{Satyalarge,Satyalarge1} is used. 
The same technique has been used successfully earlier in the field of 
random matrices 
\cite{Dyson621,Dyson622,Dyson623,Borot2012,Nadal11,Nadal11Tracy,Nadal10,Vivo07,Satyalargeright,Cunden2016,Marino2014}.

The structure of the paper is as follows: In Sec.~\ref{sec:Background} some known and relevant results of the
reduced density matrix are presented. In Sec.~\ref{sec:ResultsLeft} the large deviation function for the minimum
eigenvalue and the associated equilibrium density of states of the reduced density matrix is derived.
In Sec.~\ref{sec:ReviewResultsRight} a short review on the earlier and relevant results of the maximum eigenvalue 
is given. These results will be used in the subsequent sections.
In Sec.~\ref{sec:BipartiteEntanglement} the effect of large deviations of the extreme eigenvalues on the 
entanglement between the subsystems $A$ and $B$ is studied in detail. Then, the subsystem $A$ is divided into two 
equal parts $1$ and $2$ of dimension $N_1$ each such that $N=N_1^2$. In Sec.~\ref{sec:TripartiteEntanglement} 
the effect of these large deviations are studied on the entanglement between subsystems $1$ and $2$.

\section{STATISTICAL PROPERTIES OF THE REDUCED DENSITY MATRIX}
\label{sec:Background}

Consider the state $|\psi\rangle$ of quantum system of $A$ and $B$  
drawn from the ensemble of random pure states.
The joint probability density function (jpdf) of the 
eigenvalues  
of the reduced density matrix $\rho_A$ is then given as follows \cite{llyod,Sommers}:
\begin{eqnarray}
\begin{split}
P[\{\lambda_i\}]=&K_{M,N}\; \delta\bigg(\sum_{i=1}^N \lambda_i-1\bigg)\prod_{i=1}^N 
\lambda_i^{\frac{\beta}{2}(M-N+1)-1}\\
& \times \prod_{i<j} |\lambda_i-\lambda_j|^{\beta},
\end{split}
\label{jpdfeigen}
\end{eqnarray} 
For the $N=M$ case the jpdf corresponds to Hilbert--Schmidt measure 
whose statistical properties are well studied \cite{Sommers04}.
The normalization constant $K_{M,N}$ is calculated using the Selberg's integral 
\cite{Sommers}. 
For large $N$ and $M$, the density of the eigenvalues is given by an appropriately scaled Marcenko-Pastur (MP) 
function \cite{Nadal11,Marcenko},  
\begin{eqnarray}
\begin{split}
f(\lambda)& =  \frac{NQ}{2\pi} \frac{\sqrt{(\lambda_{+} -\lambda )(\lambda- \lambda_{-} )}}{\lambda}\\
\lambda_{\pm} &= \frac{1}{N}\bigg(1+\frac{1}{Q} \pm \frac{2}{\sqrt{Q}}\bigg), 
\end{split}
\label{MPfunct1}
\end{eqnarray}
where $\lambda\in [\lambda_{-}, \lambda_{+} ]$, $Q=M/N$ and $Nf(\lambda)d\lambda$ is the number of eigenvalues in the range 
$\lambda$ to $\lambda+d\lambda$. 
For $Q=1$ ($N=M$) the distribution has a divergence at the origin and it vanishes at $4/N$. Whereas
for $Q>1$ the eigenvalues are bounded away from zero.  

The purity of the subsystem, defined as tr$[(\rho_A)^2]$, lies between $1/N$ and $1$. For the minimum value, 
$\rho_A$ is maximally mixed and is equal to $I/N$ where $I$ is the identity matrix of dimension $N$. While for the 
maximum value the two subsystems are unentangled. The average purity of the subsystem $A$ for the random state 
$|\psi\rangle$ is given by
\begin{equation}
\left \langle \mbox{tr}\big[(\rho_A)^2\big] \right \rangle= \frac{N+M}{NM+1}\approx  \frac{1}{N}+\frac{1}{M},
\label{avepurity}
\end{equation}
where the last approximation is valid for $N,M \gg 1$ \cite{Lubkin}.  
An exact formula for the average of the von Neumann entropy is evaluated over 
the probability density in Eq.~(\ref{jpdfeigen}). It is given as follows \cite{Page,Sen,Jorge}:
\begin{eqnarray}
\label{Eq:PageFormula}
\begin{split} 
\left \langle S_{VN} \right\rangle\,&= \,\sum_{m=M+1}^{NM} \frac{1}{m}-\frac{N-1}{2M}\\
& \approx \log(N)-\frac{N}{2M}\;\; \mbox{for}\;\; 1 \ll N \leq M.
\end{split}
\end{eqnarray}
This implies that, practically there is very little information about the full pure state in a subsystem. More 
precisely, in a random pure state there is less than one-half unit of information on an average in the smaller 
subsystem of the total system.

\section{Large deviation function for the minimum eigenvalue}
\label{sec:ResultsLeft}

\begin{figure}[t!] 
\begin{center}
\includegraphics[angle=0,origin=c,scale=0.31]{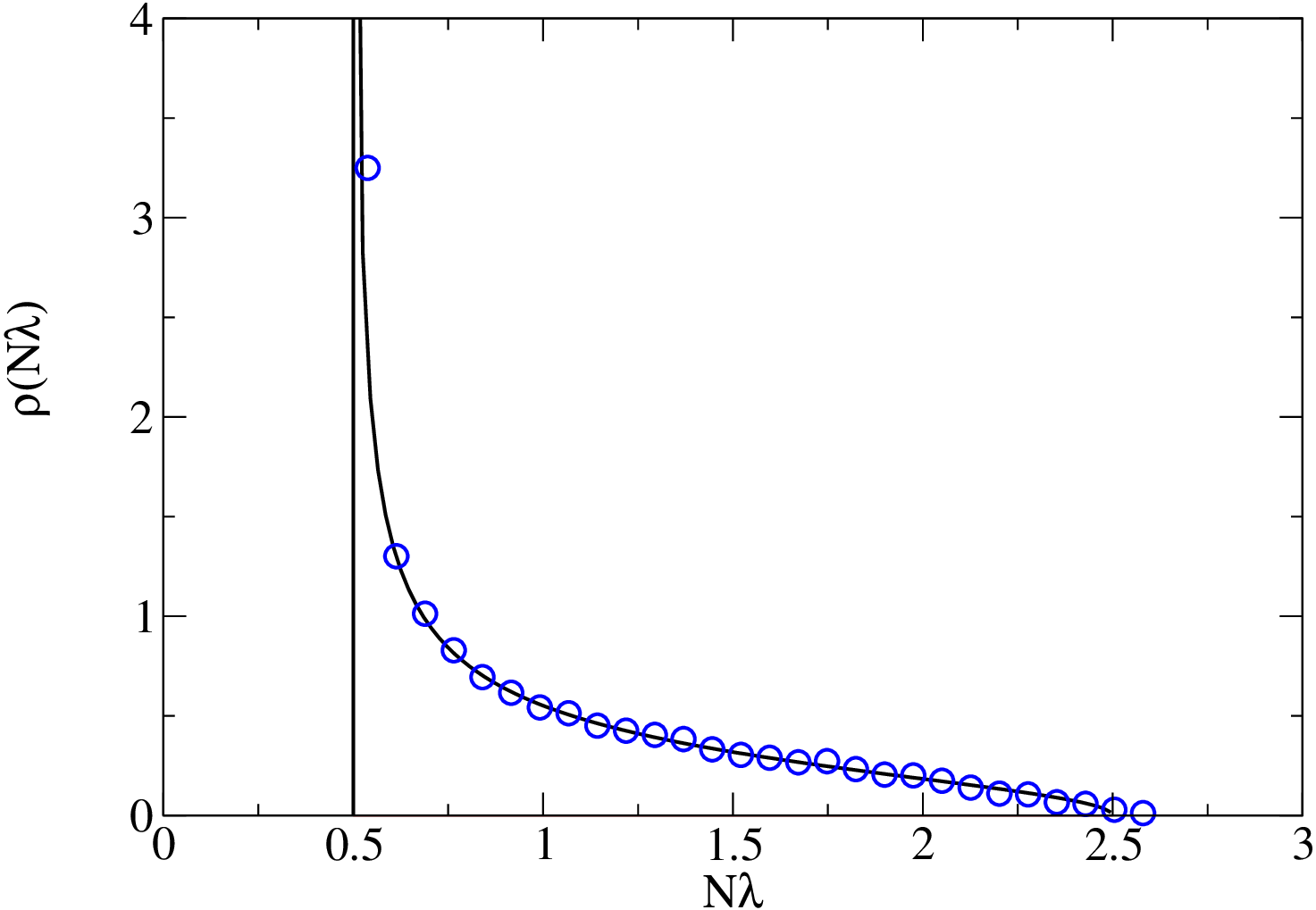} 
\caption{(Color online) Equilibrium density of the Coulomb fluid (Eq.~(\ref{Eq:zeta0p5})) when all the charges
are constrained to the right of $\zeta=0.5$ (black solid line) together with the Monte Carlo simulations 
(blue circles) for the case $N=M=100$. 
}
 \label{fig10} 
\end{center}
 \end{figure}

In this case all the rescaled eigenvalues $N\lambda$ are constrained to lie on the right side of a 
wall at $\zeta$.  
This condition is satisfied when $N\lambda_{\mbox{min}}\geq \zeta$. Since, $\lambda_{\mbox{min}}$ satisfies the 
condition $0\leq \lambda_{\mbox{min}} \leq 1/N$ which implies $0\leq \zeta \leq 1$. 
First, the results for this case will be summarized which then will be proved in Sec.~\ref{sec:ProofLeft}.
In this case the density of states of the rescaled eigenvalues for large $N$ is found as follows:
\begin{equation}
\label{Eq:LeftDensity}
 \rho({\lambda'})=\frac{1}{2\pi(1-\zeta)}\sqrt{\frac{4-3\zeta-\lambda'}{\lambda'-\zeta}},\;\;
\zeta\leq\lambda'\leq 4-3\zeta,
\end{equation}
where $\lambda'=N\lambda$ and $0\leq \zeta \leq 1$. It has a divergence at $\zeta$ and vanishes at $4-3\zeta$. 
An example for this case is demonstrated in Fig.~\ref{fig10} for the 
case of $\zeta=0.5$. For this case one obtains the following distribution:
\begin{equation}\label{Eq:zeta0p5}
\rho({\lambda'})=\frac{1}{\pi}\sqrt{\frac{5-2\lambda'}{2\lambda'-1}},\;\;\frac{1}{2}\leq \lambda' \leq \frac{5}{2}. 
\end{equation}
In the Fig.~\ref{fig10} the Monte Carlo simulations of $N=100$ is shown. It shows a good agreement 
between theory and the numerical simulations.

\begin{figure}[t!] 
\begin{center}
\includegraphics[angle=0,origin=c,scale=0.32]{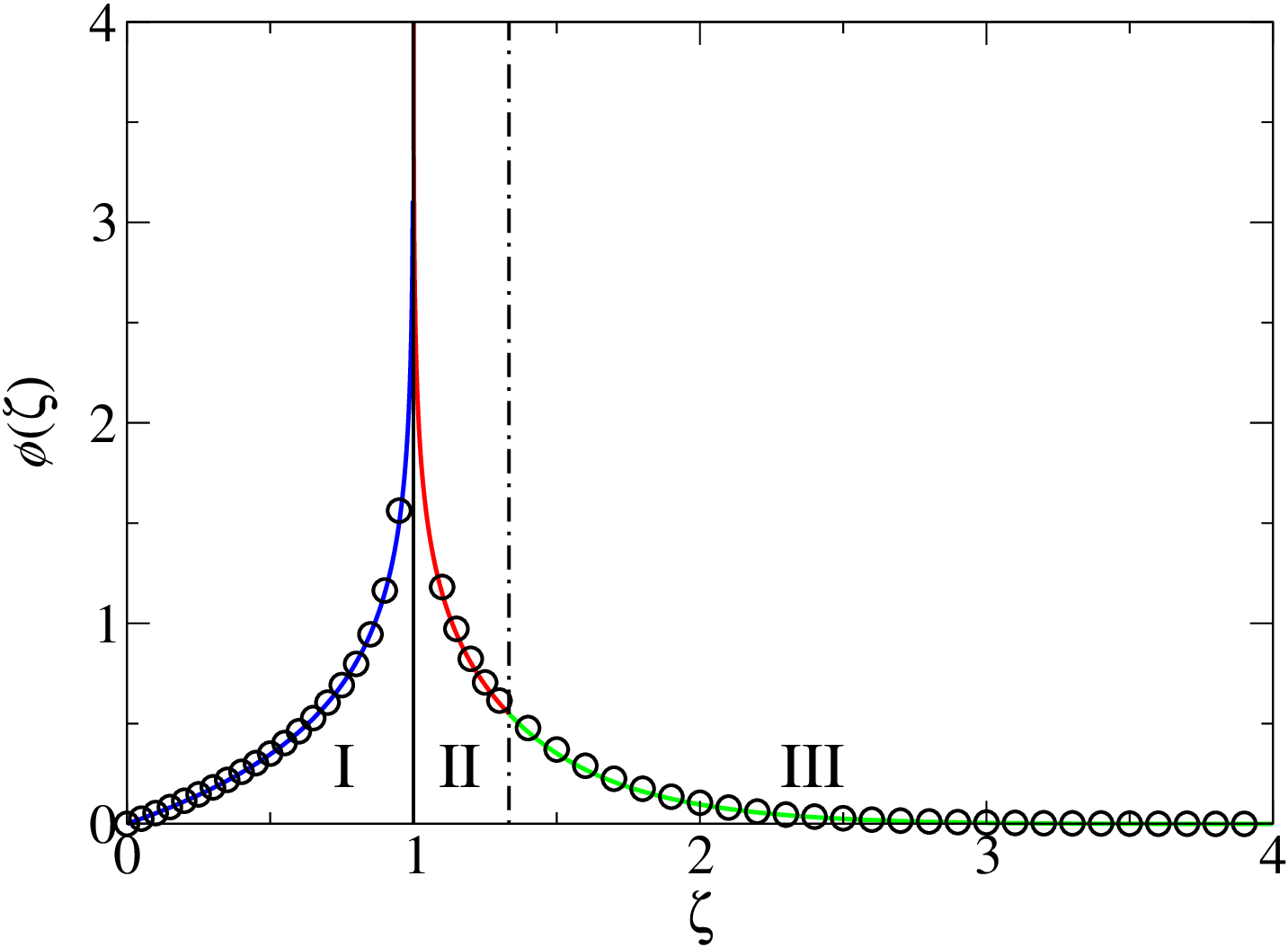} 
\includegraphics[angle=0,origin=c,scale=0.32]{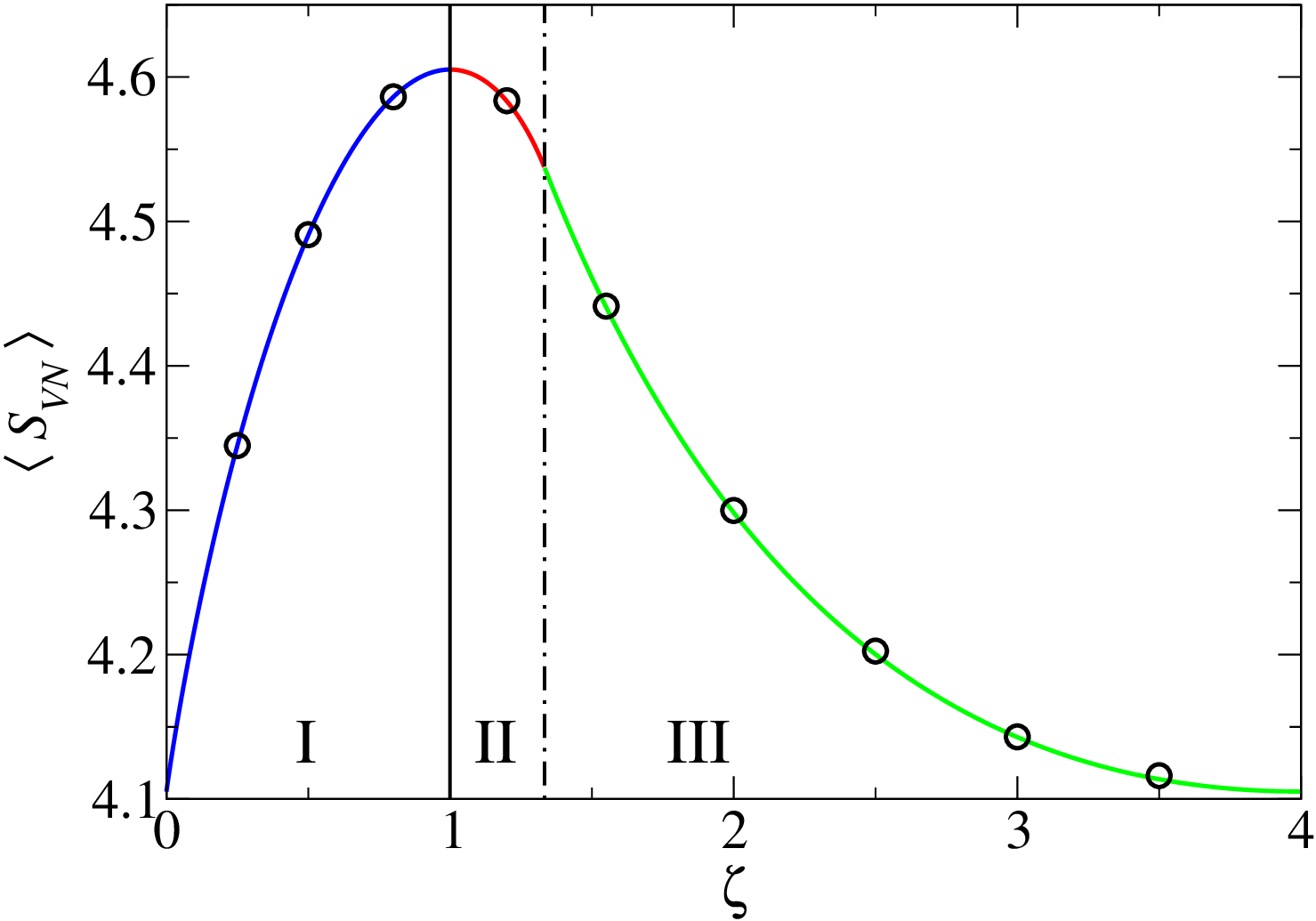} 
\caption{(Color online) Rate functions (top) and the average von Neumann entropy (bottom) as a function 
of the barrier position. Region I corresponds to $0\leq \zeta \leq 1$ when all the charges are on the right 
side of the barrier. Regions II and III corresponds to $1\leq \zeta \leq 4/3$ and $4/3 \leq \zeta \leq 4$ 
respectively when all the charges are on the left side of the barrier. Monte Carlo simulations are shown in 
both the figures using black circles for $M=N=100$. 
}
 \label{fig:RateFunction} 
\end{center}
 \end{figure}

\subsection{Evaluation of the density of states in Eq.~(\ref{Eq:LeftDensity}) using the Coulomb gas method}
\label{sec:ProofLeft}
 
The method of mapping the eigenvalues of a random matrix to a Coulomb gas problem goes back to Dyson 
\cite{Dyson621,Dyson622,Dyson623,Forresterbook}. But a major development came when the Tricomi's solution \cite{TricomiBook} was 
used first in \cite{Satyalarge,Satyalarge1} to compute the optimal charge densities and the associated
rate functions of the extreme eigenvalues of the Gaussian ensembles. This `modified Coulomb gas'
has led to a lot of developments in the field of random matrix theory and its applications 
\cite{MajumdarNumber2012,Borot2012,SatyaMajumdar2011,Castillo2010,Pierpaolo2008,SatyaIndex2009,Marino2014,Damle2011,Grabsch2017,SNMajumdar2014,Texier2013,Satyalarge1,Grabsch2016}. 
For example, problems which include finding 
the distribution of the extreme eigenvalues of the Gaussian and Wishart matrices 
\cite{Satyalarge,Satyalarge1,Vivo07,Satyalargeright,Castillo2010}, quantum transport in chaotic cavities 
\cite{Pierpaolo2008,PierpaoloVivo2010}, the index distribution for the Gaussian random fields \cite{Alan2007}
and the Gaussian ensemble \cite{SatyaIndex2009,SatyaMajumdar2011}. 
This method will be used extensively in this paper.
The definition of the rate function will be given in the subsequent part of this subsection.
The results obtained here will then be compared with the previously know in the last part of this subsection.

The unit trace constraint 
$\sum_{i=1}^N \lambda_i=1$ implies that the typical amplitude of the eigenvalues is $\lambda_{typ} \sim 1/N$.
Whereas in the case of the Wishart ensemble $\lambda_{typ}^W \sim N$. This implies that the scaling with $N$,
 for large $N$, differs in both the cases. But it should be noted that the effect of the trace constraint
does not imply the rescaling of the Wishart results by a factor of $1/N^2$. This effect of the trace constraint 
leads to a different and new behavior which includes a condensation transition, which is absent in the Wishart 
ensembles \cite{Nadal10,Nadal11,Borot2012}.

The density of states in Eq.~(\ref{Eq:LeftDensity}) corresponds to the following probability: 
\begin{eqnarray}
\label{Eq:PDFMin}
\begin{split}
P(N\lambda_{\mbox{min}}>\zeta)&=P(N\lambda_1>\zeta,N\lambda_2>\zeta,\ldots,N\lambda_N>\zeta)\\
& =\frac{\int_{\zeta}^{\infty}\ldots\int_{\zeta}^{\infty} P[\{\lambda_i\}]\prod_{i=1}^N d\lambda_i}{\int_{0}^{\infty} \ldots \int_{0}^{\infty} P[\{\lambda_i\}]  \prod_{i=1}^N d\lambda_i},
\end{split}
\end{eqnarray}
when all the eigenvalues are constrained to be larger than a fixed constant $\zeta$. The joint pdf of the eigenvalues 
$P[\{\lambda_i\}]$ given in Eq.~(\ref{jpdfeigen}) can be seen as a Boltzmann weight at inverse temperature $\beta$:
\begin{equation}
 P[\{\lambda_i\}] \propto \exp\{ -\beta E[\{\lambda_i\}]\},
\end{equation}
where the energy $E[\{\lambda_i\}]=-\gamma\sum_{i=1}^N\ln\lambda_i-\sum_{i<j}\ln|\lambda_i-\lambda_j|$
and $\gamma=1/2 -1/\beta$ (for $N=M$ case). This energy is the effective energy of a 2D Coulomb gas of charges
where the charges repel each other electrostatically via logarithmic interaction in 2D. 
For large $N$, the presence of the logarithmic interaction potential term results in the effective energy to be
of the order $E \sim O(N^2)$. 
Thus, to compute the multiple integral in Eq.~(\ref{Eq:PDFMin}) the method of steepest descent is used.
In this method, for large $N$, the configuration of $\{\lambda_i\}$ which dominates
the integral is the one that minimizes the effective energy.  
For large $N$, it can be expected that the eigenvalues are close to each other.
In that case the saddle point will be highly peaked, i.e. the most probable value and the mean will coincide.
Thus, labeling  the $\lambda_i$ by a continuous average density of states 
$\rho\left(\lambda,N \right)=N^{-1}\sum_i \langle \delta(\lambda-\lambda_i)\rangle =N \,\rho(x)$ where 
\begin{equation}\label{Eq:rhoxform}
 \rho(x)=N^{-1}\sum_i \langle\delta(x-\lambda_i N)\rangle
\end{equation} 
and $x=\lambda N$. 
Thus, the probability of $N\lambda_{\mbox{min}}$ greater than $\zeta$ can be written as 
\begin{equation}\label{pdfSigmaqContinu}
P(N\lambda_{\mbox{min}}>\zeta) \propto \int \mathcal{D}\left[\rho\right]\:
\exp\left\{-\beta N^2 \, E_\zeta\left[\rho\right]\right\}\,,
\end{equation}
where the effective energy $E_\zeta\left[\rho\right]$ is given by
\begin{eqnarray}\label{Eq:EeffSq}
&&E_\zeta\left[\rho\right]=-\frac{1}{2}\int_{\zeta}^{\infty}\int_{\zeta}^{\infty} dx \, dx' \:
\rho(x)\rho(x') \, \ln\left|x-x'\right|
\nonumber\\
&&+ \mu_0 \left(\int_{\zeta}^{\infty} dx \: \rho(x)-1\right)+
\mu_1 \left(\int_{\zeta}^{\infty} dx \:x\:  \rho(x)-1\right) 
\,.
\end{eqnarray}
The Lagrange multipliers $\mu_0$ and $\mu_1$ enforce the constraints $\int \rho(x) dx =1$ 
(the normalization of the density) and $\sum_i\lambda_i=1$ (the unit trace) respectively.
For large $N$, the method of steepest descent gives the following:
 \begin{equation} 
P\left(N\lambda_{\mbox{min}}>\zeta\right)\propto \exp\left\{-\beta N^2 
E_\zeta\left[\rho_\zeta\right]\right\}\,,
 \end{equation}
where $\rho_\zeta$ minimizes the energy (the saddle point):
\begin{equation}\label{Eq:SaddlePoint}
\frac{\delta E_\zeta}{\delta \rho}\Big|_{\rho=\rho_\zeta}=0 \,.
\end{equation}

This saddle point equation gives:
\begin{equation}\label{Eq:SaddleSq0}
\int_{\zeta}^{\infty}dx'\, \rho_\zeta(x') \ln\left|x-x'\right| = \mu_0+\mu_1 x \,.
\end{equation}
Differentiating with respect to $x$ gives:
\begin{equation}\label{Eq:SaddleSq}
\mathcal{P}\int_{\zeta}^{\infty} dx'\, \frac{\rho_\zeta(x')}{x-x'} = \mu_1 \,,
\end{equation}
where $\mathcal{P}$ denotes the Cauchy principal value.

This singular integral equation can be solved by using the Tricomi's theorem \cite{TricomiBook} which states 
that if the solution $\rho^*$ has the finite support $[L_1,L_2]$, then the finite Hilbert transform which is defined 
by the following equation  
\begin{equation}\label{Tricomi1}
F(x)=\mathcal{P}\int_{L_1}^{L_2}dx'\:\frac{\rho^*(x')}{x-x'}
\end{equation}
can be inverted as
\begin{eqnarray}\label{Tricomi}
\begin{split}
\rho^*(x)&=\frac{-1}{\pi^2 \sqrt{x-L_1}\sqrt{L_2-x}}
\bigg[C+\\
&\mathcal{P}\int_{L_1}^{L_2} dx'
 \frac{\sqrt{x'-L_1}\sqrt{L_2-x'}}{x-x'}\,F(x')\bigg]\,,
\end{split}
\end{eqnarray} 
where $C =-\pi\int_{L_1}^{L_2}dx\, \rho^*(x)$. Here, $L_1=\zeta$  and $F(x)=\mu_1$. This solution was first used 
successfully in Ref.\cite{Satyalarge,Satyalarge1} to study the large deviations of the extreme eigenvalues of 
Gaussian ensemble as mentioned in the beginning of this subsection.

The integral in Eq.~(\ref{Tricomi}) can be evaluated explicitly to obtain:
\begin{equation}
\begin{split}
\rho^*(x)&=\frac{1}{\pi \sqrt{x-\zeta}\sqrt{L_2-x}}\\
&\bigg[1+\frac{(2\left(\zeta+L_2)-4\right)(\zeta+L_2-2x)}{(\zeta-L_2)^2}\bigg],\\
&\mbox{where}\;\; \zeta \leq x \leq L_2.
\end{split} 
\end{equation}
Here, the normalization condition $\int_{\zeta}^{L_2}dx\, \rho^*(x)=1$ is used to set the constant $C=-\pi$.
Whereas $\mu_1=4(\zeta+L_2-2)/(\zeta-L_2)^2$ is obtained using the constraint $\int_{\zeta}^{L_2}dx\, x\, \rho^*(x)=1$.
There is one more unknown $L_2$ which needs to be fixed.
At the two end points $\zeta$ and $L_2$, the solution $\rho^*(x)$ either vanishes or has an inverse square root
divergence (which is integrable). When there is no constraint the density has an inverse square root
divergence at the origin and it vanishes at $4$. But when the minimum eigenvalue has to satisfy the constraint of 
being greater than $\zeta$ then intuitively it seems that the new density must have same nature at 
the boundary points as that of 
when there is no constraint. This is verified numerically for various values of $\zeta$ between zero and one.
One such illustration is shown in Fig.~\ref{fig10}.
 Thus, the condition $\rho(L_2)=0$ gives $L_2=4-3\,\zeta$. Thus, the final density as a 
function of $\zeta$ is given as follows:
\begin{equation}\label{Eq:ConstraintLeft} 
\rho(x)=\frac{1}{2\pi (1-\zeta)}\sqrt{\frac{4-3\zeta-x}{x-\zeta}},\;\; \zeta \leq x \leq 4-3\zeta. 
\end{equation}
Using $~L_2=4-3\zeta~$ the~ constant $~\mu_1~$ simplifies to $1/(2(1-\zeta))$. The constant $\mu_0$ is 
found using 
Eq.~(\ref{Eq:SaddleSq0}) and~ putting $~x=\zeta$. This gives $~\mu_0=\ln(1-\zeta)+(3\zeta-2)/(2(1-\zeta))$.
Finally, the saddle point energy is calculated. First, the saddle point Eq.~(\ref{Eq:SaddleSq0}) is multiplied by 
$\rho(x)$ then the integration is carried out. Then using Eq.~(\ref{Eq:EeffSq}) one obtains 
\begin{equation}
\label{Eq:SaddleEnergy}
 E_\zeta\left[\rho_\zeta\right]=3/4-\ln(1-\zeta)/2.
\end{equation}

Now, the rate function for the large fluctuations will be calculated. It is defined as follows. For large $N$
the probability $P\left(N\lambda_{\mbox{min}}>\zeta\right) \approx \exp\{-\beta N^2 \Phi(\zeta) \}$
where $\Phi(\zeta)$ is the rate function. The normalized probability is given as follows:
\begin{equation}\label{pdfSigmaqContinuNorm}
P\left(N\lambda_{\mbox{min}}>\zeta\right)
 \approx \frac{\int \mathcal{D}\left[\rho\right]
\:\exp\left\{-\beta N^2 \, E_\zeta\left[\rho\right]\right\}}{
\int \mathcal{D}\left[\rho\right]
\:\exp\left\{-\beta N^2 \, E\left[\rho\right]\right\}}\,,
\end{equation}
where $E_\zeta\left[\rho\right]$ is given in Eq.~\eqref{Eq:SaddleEnergy} and $E\left[\rho\right]$ is the effective energy 
associated to the joint distribution of the eigenvalues without any constraints obtained by putting $\zeta=0$ in the
Eq.~\eqref{Eq:SaddleEnergy}. Using the steepest descent method for both the numerator and the denominator 
one obtains the following:
\begin{eqnarray} \label{Eq:RateFunction}
\begin{split}
P \left(N\lambda_{\mbox{min}}>\zeta\right)
& \approx \frac{\exp\{-\beta N^2 E_\zeta[\rho_\zeta]\}}{\exp\{-\beta N^2 E[\rho^*]\}}\\
& \approx  \exp\{-\beta N^2 \Phi(\zeta)\},
\end{split}
\end{eqnarray}
with $\Phi(\zeta)=E_\zeta\left[\rho_\zeta\right]-E\left[\rho^* \right]$ and where $\rho^*$ (resp. $\rho_\zeta$) is the density 
that minimizes the energy $E\left[\rho\right]$ (resp. $E_\zeta\left[\rho\right]$). The density $\rho^*(x)$ is thus simply the 
rescaled average density of states given in Eq.~\eqref{MPfunct1} (for $Q=1$) which corresponds to $\zeta=0$ case
in Eq.~\eqref{Eq:SaddleEnergy}. Finally, the rate function is given as follows:
\begin{equation}\label{PhiNormSq}
\Phi(\zeta)_{I}=E_\zeta\left[\rho_\zeta\right]-E\left[\rho^* \right]_{\zeta=0}=-\frac{\ln(1-\zeta)}{2}\, 
\end{equation}
and is plotted in Fig.~\ref{fig:RateFunction} (region {\bf I} of top figure).
It shows a divergence as $\zeta\rightarrow 1^-$. Whereas it vanishes at $\zeta=0$ which is consistent
with the no constraint condition. The theoretically obtained curve is compared the numerically obtained 
rate function using the Monte Carlo simulations and both of them agrees very well with each other. 
The large deviations for the minimum eigenvalue of the Wishart ensemble (where there is no trace constraint) is
studied earlier in Ref.\cite{Castillo2010}. Our results, namely the rate function in the Eq.~(\ref{PhiNormSq}) 
and the density of states of the Coulomb charges differ from that of Ref.\cite{Castillo2010}. These differences 
can be attributed to the trace constraint on the reduced density matrices. 

Now, the connection of our results to the previously known ones is given.
The full distribution, for all $N=M$, of the minimum Schmidt eigenvalue and the 
minimum eigenvalue of Wishart ensemble is known \cite{RMTBook}. An exact relation between the two is also known
\cite{RMTBook}. These results have been evaluated for the three cases of $\beta=1$, $2$ and $4$.
From these earlier results it can be seen easily that the large deviation tail is 
$(1-\zeta)^{(\beta N^2/2)}$ which gives the rate function as $-(1/2)\ln(1-\zeta)$. This expression
 agrees with our expression for $\Phi(\zeta)_{I}$.  
But our calculations, using the Coulomb gas method from \cite{Satyalarge,Satyalarge1}, shows that this expression 
for the rate function holds for all the values of $\beta$ and not just for these three values.
The connections of the rate function $\Phi(\zeta)_{I}$ and the corresponding equilibrium density in 
Eq.~(\ref{Eq:ConstraintLeft}) to those of maximum eigenvalue will be given in the next section.

\section{Review of results of the maximum eigenvalue}
\label{sec:ReviewResultsRight}

In this section, a short review on the relevant results of the maximum eigenvalue will be given. These results 
will be used 
in the subsequent parts of this paper. The question addressing the constraint on all the eigenvalues being less 
than a fixed constant $\zeta$ is studied in detail earlier in \cite{Nadal11} using the modified Coulomb gas method 
from Ref.\cite{Satyalarge,Satyalarge1}. This constraint is equivalent to the condition that $\lambda_{\mbox{max}}\leq\zeta$.
For equal dimensionality $N=M$ case i.e. $Q=1$ the rescaled eigenvalues lie in the interval $(0,4]$ 
(refer Eq.~(\ref{MPfunct1})). Thus, the barrier position $\zeta$ is effective only when $\zeta\leq 4$. Since, 
$\lambda_{\mbox{max}}$ satisfies the condition $1/N\leq\lambda{\mbox{max}}\leq1$ which implies $1\leq \zeta \leq 4$. 
Throughout this paper, whenever $\zeta$ lies between zero (one) and one (four) it refers to the fact 
$N\lambda_{\mbox{min}}\geq \zeta$ ($N\lambda_{\mbox{max}}\leq \zeta$).

In Ref.\cite{Nadal11} it was shown that there are two regions depending on the nature of the density which shows a 
transition at $\zeta=4/3$. Thus, there are two sub cases. Case one ($4/3 \leq \zeta \leq 4$): the density has a 
support on $[0,\zeta]$ and has a divergence at both the boundaries except at $\zeta=4/3$ where the density vanishes
at the origin. Case two ($1\leq \zeta \leq 4/3$): the density has a support on $[4-3\zeta,\zeta]$. In this case it 
vanishes at $4-3\zeta$ and has a divergence at $\zeta$. 
The density in the first case is given as 
\begin{equation}
\rho(x)=\frac{2\zeta^2+4(\zeta-2)(\zeta-2 x)}{2\pi\zeta^2\sqrt{x(\zeta-x)}},\;\;
0\leq x\leq\zeta;
\label{Eq:ConstraintRight}
\end{equation}
whereas in the second case it is given as 
\beq
\label{Eq:ConstraintRight2}
\rho(x)=\frac{1}{2\pi(\zeta-1)} \sqrt{\frac{3\zeta-4+x}{\zeta-x}}, \;\; 4-3\zeta\leq x\leq \zeta.\\
\eeq
The rate functions were also derived in \cite{Nadal11}. For the first case it is given as
\begin{equation}
\Phi(\zeta)_{III}=\dfrac{3}{4}-4\dfrac{\zeta-1}{\zeta^2}-\dfrac{1}{2}\ln\left(\dfrac{\zeta}{4}\right)
\end{equation}
and is plotted in Fig.~\ref{fig:RateFunction} (region {\bf III} of the top figure). It vanishes at $\zeta=4$ which 
is consistent with the no constraint condition. Whereas for the second case it is given as
\begin{equation}
 \Phi(\zeta)_{II}=-\dfrac{\ln(\zeta-1)}{2}
\end{equation}
and is plotted in Fig.~\ref{fig:RateFunction} (region {\bf II} of top figure). 
It can be seen from Eq.~(\ref{PhiNormSq}) that the rate functions $\Phi(\zeta)_{I}$ (derived in 
Sec.~\ref{sec:ResultsLeft} of this paper) and $\Phi(\zeta)_{II}$ are reflections of each other around $\zeta=1$, in fact it can 
be seen from the densities in Eqs.~(\ref{Eq:ConstraintRight2}) and (\ref{Eq:LeftDensity}) that both are reflections 
of each other around $\zeta=1$ provided $2/3\leq\zeta\leq1$ is used for the deviations of the minimum eigenvalue.

The large deviations for the maximum eigenvalue of the Wishart ensemble (where there is no trace constraint) is 
studied in Ref.\cite{Vivo07}. But, no transition in the density as well as the rate function is observed there, which 
can be attributed to the absence of the trace constraint on the matrices. For the $N=M$ case in Ref.\cite{Vivo07} 
the density shows divergence at both the ends of its eigenvalue support whenever $\lambda_{\mbox{max}}<4$.

\section{Bipartite entanglement}
\label{sec:BipartiteEntanglement}

In the earlier works in Ref.\cite{Nadal10,Nadal11}, using the Coulomb gas method the full probability 
distribution of the Renyi entropy, a 
measure of bipartite entanglement of which von Neumann entropy is a special case, is derived. 
There, two critical points are found for which the charge density shows a transition as the value of Renyi entropy 
is varied. In the first transition, the integrable singularity at the origin disappears while in the second, the 
largest eigenvalue gets detached from the continuum sea of all the other eigenvalues.  
As explained in Sec.\ref{sec:ReviewResultsRight} the density
shows a transition when there are large deviations in the maximum eigenvalue while no such transition is observed 
for the same in the minimum eigenvalue. 

In the introduction of this paper importance of the extreme eigenvalues from the perspective of entanglement 
between subsystems $A$ and $B$ is given. But, what is the actual entanglement measured using the von Neumann 
entropy for given constraints (case of large deviations here) on these extreme eigenvalues is unanswered.
Thus, in this section our aim is to quantify the 
bipartite entanglement between subsystems $A$ and $B$ when there are large deviations in the extreme eigenvalues 
from their average values. We would also like to investigate that does the signature of presence or absence of 
transition in the densities are reflected in the entropies or not.
Here, the von Neumann entropy is used as a measure of entanglement 
\cite{Bennett96,Zyczkowski06}. For this study the optimal Coulomb charge densities obtained in 
Sec.\ref{sec:ResultsLeft} and the ones from earlier studies reviewed in Sec.\ref{sec:ReviewResultsRight} will be used. 
It should be noted that when there no large deviations in the extreme eigenvalues the average von Neumann entropy
is known as Page's formula \cite{Page} and is given in Eq.(\ref{Eq:PageFormula}).
Thus, with this conditional average generalization of this formula will also be addressed.
Results obtained in this section and those on tripartite entanglement in the next section are compared qualitatively
from the perspective of monogamous nature of entanglement at the end of next section.

As a first case, the large deviations in the case of maximum eigenvalue is considered. As pointed out 
in the earlier parts of this paper, there are two sub cases depending on the position of the barrier.
Using Eq.~(\ref{Eq:vonNeumannEntropy}) and labeling the eigenvalues of $\rho_A$ by a continuous average density
of states as done in Sec.~\ref{sec:ProofLeft} the average von Neumann entropy is given as
\begin{equation}\label{Eq:AveVonNeumann}
\langle S_{VN} \rangle = -N\int x \ln(x) \rho(x) dx
\end{equation}
where the form of $\rho(x)$ is given in Eq.~(\ref{Eq:rhoxform}). One needs to use the appropriate expression of the
charge density
$\rho(x)$ depending on the value of $\zeta$ for calculating the average entropy. For the case when 
$4/3 \leq \zeta \leq 4$ 
the density given in the Eq.~(\ref{Eq:ConstraintRight}) is used in Eq.~(\ref{Eq:AveVonNeumann}) to calculate the
average entropy. Then using {\it Mathematica~9} it is found to be 
\begin{eqnarray}\label{Eq:Von4by3to4}
\ln\left(\frac{4N}{\zeta}\right)+\frac{\zeta}{4}-\frac{3}{2}.
\end{eqnarray}
It is plotted in Fig.~\ref{fig:RateFunction} for the case $N=100$ (region {\bf III} of the bottom figure). 
For the special case of $\zeta=4$ which corresponds to no constraint on the maximum eigenvalue, the average von 
Neumann entropy turns out to be $\ln(N)-1/2$ \cite{Page}. This value agrees very well with that derived in 
Ref.\cite{Page} where there are no additional constraints on the eigenvalues of $\rho_A$.

For the second case when $1 \leq \zeta \leq 4/3$ the density given in the Eq.~(\ref{Eq:ConstraintRight2}) 
is used in Eq.~(\ref{Eq:AveVonNeumann}) to obtain the average von Neumann entropy. 
Again using {\it Mathematica~9} it turns as follows:
\begin{eqnarray}\label{Eq:Von1to4by3}
&& \ln(N)-\frac{1}{\zeta(4-3\zeta)}\bigg((\zeta-a)^2(3\zeta-4)\times \nonumber \\
&& _pF_q\left[\{1,1,3/2\},\{3,4\},\frac{4(\zeta-1)}{3\zeta-4}\right] + \nonumber \\
&& 2(\zeta-1)^2(9\zeta-10)_pF_q\left[\{1,1,5/2\},\{3,4\},\frac{4(\zeta-1)}{3\zeta-4}\right]\nonumber \\
&& -(3\zeta-4)\left(8-19\zeta+11\zeta^2+\zeta\ln\left(4-3\zeta\right)\right)
\bigg)
\end{eqnarray}
where $_pF_q\left[ a,b,z \right]$ is the generalized hypergeometric function. It is plotted in 
Fig.~\ref{fig:RateFunction} for $N=100$ (region {\bf II} of the bottom figure). The special case when $\zeta=1$ 
is now considered. In that case the maximum eigenvalue is equal to $1/N$
which implies the von Neumann entropy is $\ln(N)$ which is also the maximum value it can take
as explained in the introduction of the paper.  
It can also be evaluated using the Eq.~(\ref{Eq:Von1to4by3}). The entropy for $N=100$ 
indeed equals $\ln(100)\approx 4.605$. The Eqs.~(\ref{Eq:Von4by3to4}) and (\ref{Eq:Von1to4by3})
are compared with the Monte Carlo simulations as shown in bottom figure of Fig.~\ref{fig:RateFunction}.
It can be seen that the numerical simulations agrees very well with the analytical results.

At $\zeta=4/3$ (the transition between regimes {\bf II} and {\bf III}), the average von Neumann entropy 
$\langle S_{VN} \rangle$ has a 
nonanalyticity. It is continuous with $\langle S_{VN} \rangle(4/3)=\ln(3N)-7/6$ and once differentiable
with $\frac{d\langle S_{VN} \rangle}{d\zeta}\big|_{\zeta=4/3}=-1$.
However, the second derivative is discontinuous: 
$\frac{d^2\langle S_{VN} \rangle}{d\zeta^2}\big|_{\zeta=4/3^-}=-9/2$ but
$\frac{d^2\langle S_{VN} \rangle}{d\zeta^2}\big|_{\zeta=4/3^+}=9/16$. Thus, similar to rate function the von 
Neumann entropy shows a discontinuity but in its second derivative at $\zeta=4/3$.
Thus, the signature of the transition in the density of states can be observed in the von Neumann entropy.

Now, the case of the large deviations of the minimum eigenvalue is considered. 
The barrier position $\zeta$ satisfies $0\leq\zeta\leq1$. Computing analytical expression
for the entropy is difficult. Thus, it is evaluated numerically using the density in Eq.~(\ref{Eq:ConstraintLeft}) 
and Eq.~(\ref{Eq:AveVonNeumann}) for the case of $N=100$. It is plotted in Fig.~\ref{fig:RateFunction} 
(region {\bf I} of the bottom figure) along with the Monte Carlo simulations.
It can be seen that both agree with each other very well. 
It can be seen easily from the figure the entropy is continuous and infinitely differentiable in the 
region {\bf I} since it is concave downward.
This can be attributed to the fact that density don't show any transition in this case.

\section{Entanglement within subsystems}
\label{sec:TripartiteEntanglement}

In this section, the subsystem $A$ is further divided into two parts denoted as $1$ and $2$ having Hilbert space 
dimension $N_1$ and $N_2$ respectively such that $N=N_1N_2$. Then the effect of the large deviations of the extreme 
eigenvalues of $\rho_A$ are studied on the entanglement between the subsystems $1$ and $2$. Now, we have a 
tripartite pure state having dimensions $N_1$, $N_2$ and $M$. The entanglement in such a tripartite pure system 
when its state is chosen randomly, has been studied previously in Ref.\cite{Uday12,Aubrun12,Aubrun2014}. There it is shown 
that the entanglement between subsystems $1$ and $2$ shows a transition at $M=4N_1N_2$ for sufficiently large 
subsystem dimensions.  
 
The entanglement between subsystems $1$ and $2$ is studied using the log negativity measure \cite{vidal}.
It is defined as $E_{LN}(\rho_{12}) = \log(||\rho_{12}^{\Gamma}||)$, where $||\rho^{\Gamma}||$ is the trace norm of 
the partial transpose (PT) matrix $\rho^{\Gamma}$~\cite{peres96}. When the log negativity is greater than zero,
the state is said to have the negative partial transpose (NPT). Then the state is entangled. 
When the log  negativity is zero, the state is said to have the positive partial transpose (PPT).
Then the state is either separable or bound entangled \cite{mhorodeckibound}.

Now, the numerical procedure to generate random states $\rho_{12}$ having large deviations in their extreme 
eigenvalues is given. Every density matrix, which is Hermitian, can be diagonalized by an unitary rotation $U$. 
It is thus natural that the distribution of eigenvalues and that of eigenvectors of $\rho_{12}$ are independent. 
Thus, the probability measure of $\rho_{12}$ factorizes in a product form \cite{Sommers,Hall1998}, 
$\mbox{d}\mu_x=\mbox{d}\nu_x(\lambda_1,\lambda_2,\ldots,\lambda_N) \times \mbox{d}h$.
Here, $\lambda_1,\lambda_2,\ldots,\lambda_N$ are the eigenvalues of $\rho_{12}$ and the factor $ \mbox{d}h$ 
determines the distribution of its eigenvectors. The probability measure used for the eigenvalues is given in 
Eq.~(\ref{jpdfeigen}) along with the constraint on the extreme eigenvalues. For the measure  $ \mbox{d}h$ the unique Haar 
measure on $U(N)$ is taken which determines the statistical properties of the eigenvectors forming $U$. Thus, this 
gives $\rho_{12}=Ud\,U^{\dagger}$ where $d$ is a diagonal matrix $[\lambda_1,\lambda_2,\ldots,\lambda_N]$.
The eigenvalues are generated numerically using the Monte Carlo method. Whereas the matrix $U$ is generated using 
the algorithm given in Ref.\cite{Mezzadri2007}.

Earlier works have studied the effect of PT on $\rho_{12}$ in tripartite random pure states \cite{Uday12,Aubrun12,Aubrun2014}.
It is shown that the density of $\rho_{12}$ after PT is very close to the Wigner's semicircle law 
when the dimensions of both the subsystems are not too small and are of the same order. 
In fact, the Wigner's semicircle law is also obtained in a bipartite mixed state $\rho_{12}$ after PT 
which is obtained by uniformly mixing sufficiently large number of random bipartite pure states \cite{Marko}. 
But in this paper our focus is on the bipartite and tripartite random pure system.
In these works the extreme eigenvalues fluctuates around their average values. In Ref.\cite{Uday12} the minimum 
eigenvalue of $\rho_{12}^{\Gamma}$ is shown to follow the Tracy-Widom distribution. Using this the fraction of 
entangled states at criticality ($M=4N_1N_2$) was given.  This suggests to investigate the effects of the 
large deviations of 
the extreme eigenvalues of $\rho_{12}$ on the density of $\rho_{12}^{\Gamma}$ as well as on the entanglement 
between subsystems $1$ and $2$. 
The results are plotted in Figs.~\ref{fig:PTzetaAll}, \ref{fig:PTzetaAllRight} and \ref{fig:negativity}
for the case $N_1 = N_2=10$ and $M=100$.  
It can be seen that the eigenvalue densities of $\rho_{12}^{\Gamma}$ is very close to the Wigner semicircle law. 
As the barrier position is changed an entanglement transition takes place from dominantly NPT states to dominantly 
PPT.

It should be mentioned here that the case $N_1\neq N_2$ without any large deviations in the extreme eigenvalues
has been studied in \cite{Uday12}. There it was shown that the density of states of $\rho_{12}^{\Gamma}$
had a skewness which was calculated analytically. It is also observed in our work that the density has a skewness
(not presented here) but calculating it analytically seems to be mathematically challenging. Thus, it is not 
addressed in this paper.

\subsection{Model for shifted semicircles}

In the earlier work in Ref.\cite{Uday12} the semicircular density of $\rho_{12}^{\Gamma}$ was well studied using a 
simple model. It was suggested by using the fact that the first two moments remains unchanged under the PT operation.
The semicircular density depends only on two moments, the mean and the variance. Thus, it was proposed 
to shift and scale the semicircle of the Gaussian ensembles such that the first two moments of 
$\rho_{12}$ are matched. 
To explain the semicircular density obtained in this paper the same model from Ref.\cite{Uday12} is used. The model 
has been used to accurately predict the transition from the dominantly NPT states to the dominantly PPT states.

Now, the model for the shifted semicircles is given. Here, it is assumed that these random matrices 
are sampled from the Gaussian unitary ensemble (GUE). Thus, consider
\beq
Y=X+\dfrac{I_N}{N}
\label{model}
\eeq
where $X$ is a $N\times N$ random matrix from the GUE ensemble with the necessary matrix element variance 
such that it matches with that of $\rho_{12}$, and $I_N$ is the identity matrix of dimension $N$. 
It can be seen that 
$\langle \mbox{tr}(Y)\rangle =1$ since $\langle \mbox{tr}(X)\rangle =0$, where the angular brackets indicates 
the ensemble average. Here, the case of large matrix dimension is considered. Thus, it can be expected that the 
influence of the fact that the $\mbox{tr}(Y)$ is not exactly equal to one for each and every member of the 
ensemble will not be observed except in the case of very small dimensional cases.

\begin{figure}[t!] 
\begin{center}
\includegraphics[angle=0,origin=c,scale=0.31]{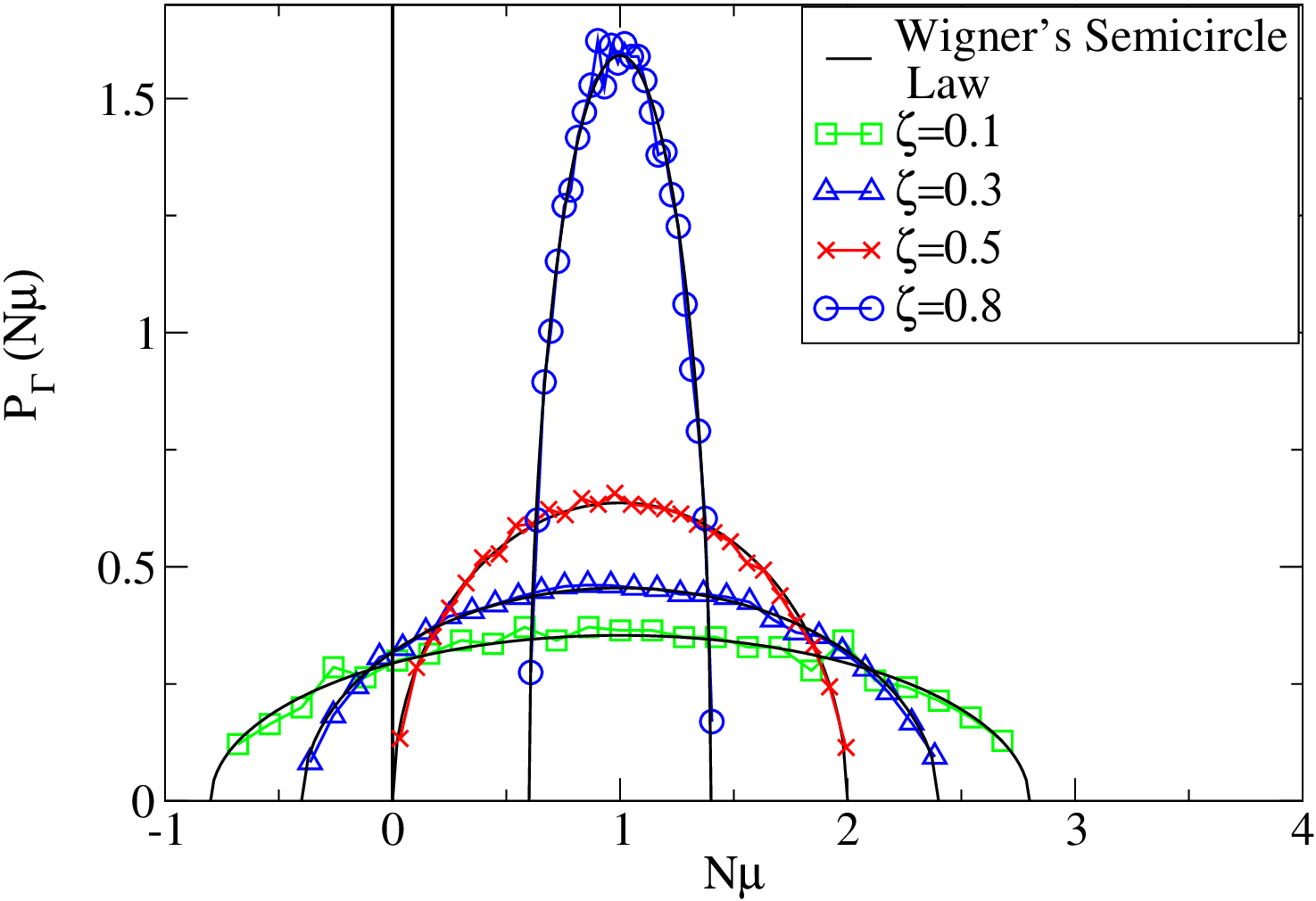} 
\caption{(Color online) 
Density of states of $\rho_{12}^{\Gamma}$ for various values of barrier positions $\zeta$ between zero and one. 
All the eigenvalues of randomly chosen $\rho_{12}$ are greater than the barrier position.
One thousand such matrices are used for each $\zeta$. 
It corresponds to the large deviations of the minimum eigenvalue $(0\leq \zeta \leq 1)$.  
A vertical line at the origin has been shown to draw attention to the negative part of the spectrum.
Here, $N_1=N_2=10$ and $M=100$. 
}
\label{fig:PTzetaAll} 
\end{center}
\end{figure}

\begin{figure}[t!] 
\begin{center}
\includegraphics[angle=0,origin=c,scale=0.31]{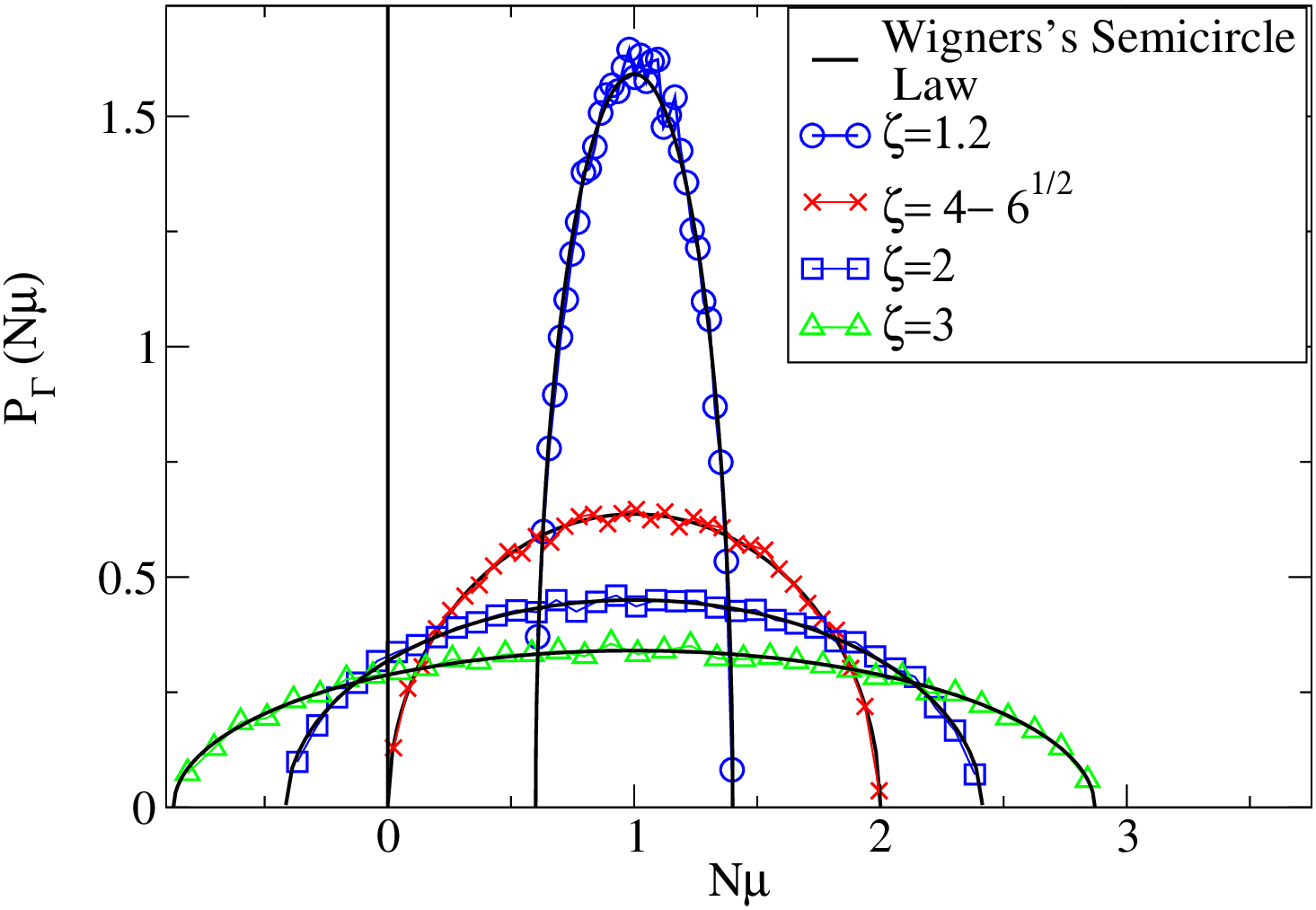} 
\caption{(Color online) 
Density of states of $\rho_{12}^{\Gamma}$ for various values of barrier positions $\zeta$ between one and four. 
All the eigenvalues of randomly chosen $\rho_{12}$ are smaller than the barrier position. One thousand such matrices 
are used for each $\zeta$. It corresponds to the large deviations of the maximum eigenvalue $(1 \leq \zeta \leq 4)$. 
A vertical line at the origin has been shown to draw attention to the negative part of the spectrum.
Here, $N_1=N_2=10$ and $M=100$. 
}
\label{fig:PTzetaAllRight} 
\end{center}
\end{figure}

It can be seen that the eigenvalues of $Y$ are all those of $X$ shifted by $1/N$. Thus, considering the 
spectrum of $X$ alone will be sufficient. 
Under the assumption that $X$ is sampled from the GUE it follows that the density of eigenvalues of $Y$ for large $N$
is given as follows:
\beq
 P(\mu)=\frac{2}{\pi R^2}\sqrt{R^2-\bigg(\mu-\frac{1}{N}\bigg)^2},\; -R+\frac{1}{N} < \mu < R+\frac{1}{N},
\label{wignerdist}
\eeq
where 
\begin{equation}
\label{Eq:Rform}
R=2\sqrt{\dfrac{1}{N} \langle\mbox{tr}(X^2)\rangle}= 2 \sqrt{\frac{1}{N}\br \tr(\rho_{12}^2) \kt \, -\, \frac{1}{N^2}}.
\end{equation} 
Now, the scaled variable $x= \mu N$ is used. This results into the semicircular probability density having a shift 
of $1$ and a rescaled ``radius" $\tilde{R} = N R$. Explicitly:
\beq
P_{\Gamma}(x)=\frac{2}{\pi \tilde{R}^2}\sqrt{\tilde{R}^2-(x-1)^2},\; \;1-\tilde{R} < x < 1+\tilde{R}.
\label{wignerdistscaled}
\eeq
This is the the Wigner semicircle law that has been observed in Figs.~(\ref{fig:PTzetaAll}) and 
(\ref{fig:PTzetaAllRight}).
Now, $\tilde{R}$ is calculated when there are large deviations in the extreme eigenvalues. This requires to find the average 
purity of $\rho_{12}$. First, the case of large deviations of the minimum eigenvalue is considered. Using
the density of states in Eq.~(\ref{Eq:LeftDensity}) in {\it Mathematica~9}, the purity turns out to be 
$\langle\mbox{tr}(\rho_{12}^2)\rangle=P_1/N=(2-2\zeta+\zeta^2)/N$. This gives the rescaled radius 
$\tilde{R}=2(1-\zeta)$ where $0\leq \zeta \leq 1$. 
Similarly, for the case of the large deviations of the maximum eigenvalue, the density of states in 
Eqs.~(\ref{Eq:ConstraintRight2}) and (\ref{Eq:ConstraintRight}) is used. The purity is found to be 
$P_2/N=(2-2\zeta+\zeta^2)/N$ and $P_3/N=-\zeta (\zeta-8)/(8N)$ for $1 \leq \zeta \leq 4/3$ and 
$4/3\leq \zeta \leq 4$ respectively. Here, $P_1$, $P_2$ and $P_3$ are the rescaled purities. Using these purities 
$\tilde{R}$ equals  
$2(\zeta-1)$ and $2\sqrt{(-\zeta^2+8\zeta-8)/8}$ for $1 \leq \zeta \leq 4/3$ and $4/3\leq \zeta \leq 4$ respectively. 
It can be seen that these analytical expressions
for the rescaled radii agrees very well with those from Figs.~(\ref{fig:PTzetaAll}) and (\ref{fig:PTzetaAllRight}). 

This model gives the NPT-PPT transition very well. It can be seen that the condition for this transition
is $\tilde{R}=1$. Using this condition one obtains $\zeta=1/2$ and $4-\sqrt{6}$ as the transition points for the 
large deviations
of the minimum and maximum eigenvalue respectively. For any $\zeta>1/2$ in the case of minimum eigenvalue and 
$\zeta<4-\sqrt{6}$ in the case of maximum eigenvalue the radius is smaller than one and there are predominantly 
PPT states. Whereas in the opposite cases the lower bounds are such that there are predominantly NPT states. Thus, this 
simple model from Ref.\cite{Uday12} of a shifted random matrix of the GUE kind for the partial transpose gives the 
transition very well. 
These critical values of the barrier positions can be 
observed in Figs.~(\ref{fig:PTzetaAll}) and (\ref{fig:PTzetaAllRight}).

In Ref.\cite{Uday12} it was shown analytically that before and after the PT the range of the eigenvalues is the same.
Extreme deviations from this result were shown to occur when the state $\rho_{12}$ is pure or nearly pure.
For the large deviation of the minimum eigenvalue the density before PT has a support on $[\zeta,4-3\zeta ]$
and after PT it becomes $[1-\tilde{R},1+\tilde{R}]$ where $\tilde{R}=2(1-\zeta)$ where $0\leq\zeta\leq1$. Thus, 
the range of the eigenvalues before and after the PT are both equal to $4(1-\zeta)$.

Similarly, for the large deviations of the maximum eigenvalue the density before PT has a support on 
$[ 4-3\zeta,\zeta ]$ and $[ 0 ,\zeta ]$ for $1 \leq \zeta \leq 4/3$ and $4/3\leq \zeta \leq 4$ respectively.
After PT the support is again $[1-\tilde{R},1+\tilde{R}]$ but with $\tilde{R}=2(\zeta-1)$ and 
$2\sqrt{(-\zeta^2+8\zeta-8)/8}$ for $1 \leq \zeta \leq 4/3$ and $4/3\leq \zeta \leq 4$ respectively. 
Thus, it can be seen that only for $1 \leq \zeta \leq 4/3$ the range of the eigenvalues before and after PT equals 
$4(\zeta-1)$. This range is reflection symmetry of that corresponding to the large deviations 
of the minimum eigenvalue around $\zeta=1$.
While for $4/3\leq \zeta \leq 4$ the range of the eigenvalues after PT is larger than that of before PT
except at $\zeta=4/3$ and $4$ where both the ranges are equal. It should be mentioned that these results are 
valid for the case $N_1\neq N_2$ since they depend only on $N=N_1N_2$ and $M$.
But when $N_1$ and $N_2$ differ significantly the density of states of $\rho_{12}^{\Gamma}$ has a skewness
whereas the model predicts zero skewness.

\subsection{Logarithmic negativity}

The average log negativity between two subsystems 1 and 2 is now studied. 
The formalism from Ref.\cite{Uday12} is used again where the fact that the density of states 
after PT is Wigner's semicircle was used. 
There it is shown analytically that 
\beq
\br E_{LN}\kt_M =   \log \Bigg[\frac{2}{\pi} \sin^{-1}\Big(\frac{1}{\tilde{R}}\Big)+
      \frac{2}{3 \pi \tilde{R}}\sqrt{1-\frac{1}{\tilde{R}^2}} \left(1+2 \tilde{R}^2\right) \Bigg].  
\label{lognegform}
\eeq

\begin{figure}[t!] 
\begin{center}
\includegraphics[angle=0,origin=c,scale=0.31]{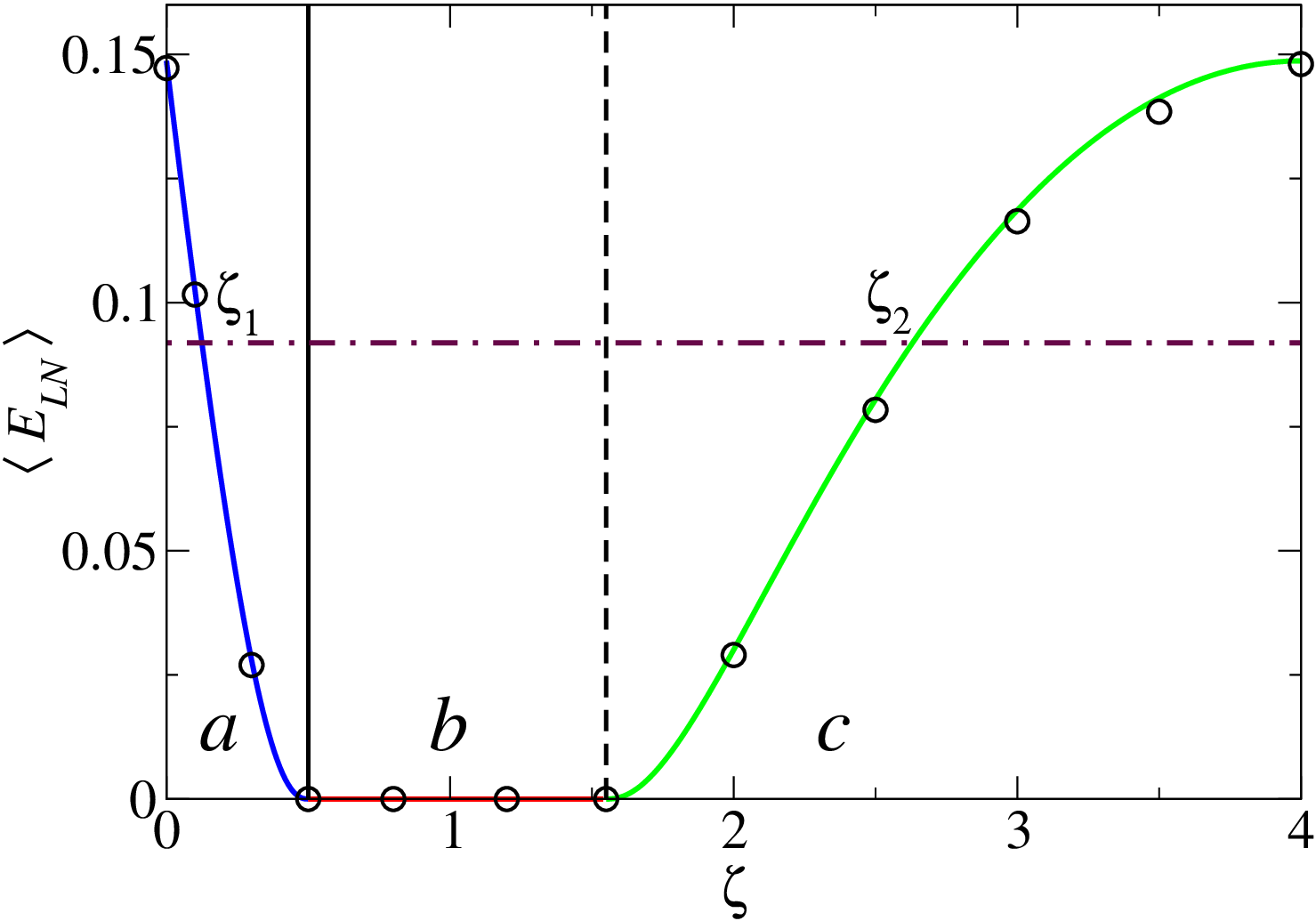} 
\caption{(Color online) Average entanglement in random states $\rho_{12}$ as measured by the log negativity between 
subsystems $1$ and $2$ for various barrier positions. This is compared with the analytical result in
Eq.~(\ref{lognegform}) based on the simple model. Black solid vertical line (dotted line) corresponds to $\zeta=1/2$  
($\zeta=4-\sqrt{6}$) showing the entanglement transition due to large deviations of the minimum (maximum) eigenvalue. 
Horizontal dash-dotted line is drawn such that $\zeta$ between $0$ and $1$ corresponding to its intersection 
with the log negativity is $1/8$.  
Regions $a$, $b$ and $c$ corresponds to $0\leq\zeta\leq1/2$, $1/2\leq\zeta\leq 4-\sqrt{6}$ and 
$4-\sqrt{6}\leq\zeta\leq 4$ respectively. Here, $N_1=N_2=10$ and $M=100$.
}
\label{fig:negativity} 
\end{center}
\end{figure}

Here, $\br E_{LN}\kt_M$ denotes the log negativity obtained using the simple model. This formula is valid only 
for $\tilde{R}\geq 1$ otherwise $\br E_{LN}\kt_M$ is zero. For our case,
$\tilde{R} = NR= 2(1-\zeta)$ $(0\leq\zeta\leq 1)$ for the large deviations of the minimum eigenvalue. 
Whereas, $\tilde{R}$ is $2(\zeta-1)$ and $2\sqrt{(-\zeta^2+8\zeta-8)/8}$ for $1 \leq \zeta \leq 4/3$ and 
$4/3\leq \zeta \leq 4$ respectively for the large deviations of the maximum eigenvalue.
 For the critical case $\tilde{R}=1$ this formula gives zero for the 
average log negativity. When $\tilde{R}<1$, the states obtained are predominantly PPT. In that case 
$\br E_{LN}\kt =0$. Thus, it can be seen that $\br E_{LN}\kt =0$ for $1/2\leq\zeta\leq4-\sqrt{6}$ since 
$\tilde{R}\leq1$ for this range of $\zeta$ as shown in the previous subsection.  
The Eq.~(\ref{lognegform}) is plotted in Fig.~\ref{fig:negativity} along with 
numerical results for various values of $\zeta$ for $N_1=N_2=10$ and $M=100$. 
It can be seen that Eq.~(\ref{lognegform}) works very well.
Consider the situations in which there are no constraints on either of the extreme eigenvalues.
It implies $\zeta=0$ ($\zeta=4$) for the minimum (maximum) eigenvalue. This gives $\tilde{R}=2$ for both of them.
In that case the Eq.~(\ref{lognegform}) gives $\br E_{LN}\kt \approx 0.148702$. 
This value can be observed in Fig.~\ref{fig:negativity} at $\zeta=0$ and $\zeta=4$.

Another interesting features that is observed in Fig.~\ref{fig:negativity} is that there are two different values
of $\zeta$'s ($\zeta_1$ and $\zeta_2$, say) corresponding to the large deviations of the extremes for which 
entanglement between subsystems $1$ and $2$ is same. Here, $\zeta_1$ ($\zeta_2$) corresponds to the large deviation 
of the minimum (maximum) 
eigenvalue. Thus, this implies $0\leq\zeta_1\leq1$ and $1\leq\zeta_2\leq4$. It can be seen 
that from Eq.~(\ref{lognegform}) for the log negativity, derived using the simple random matrix model, that two 
different $\zeta$'s will 
result in the same log negativity provided $\tilde{R}$ is same for both of them. Whereas in Eq.~(\ref{Eq:Rform}) it 
is shown that $\tilde{R}$ depends only on the purity of $\rho_{12}$. 
Thus, this implies that large deviations of the extremes will have 
the same log negativity if the corresponding purities (so does the rescaled purities) are same.

Using the simple model it is shown that log negativity is non-zero when $\zeta<1/2$ ($\zeta>4-\sqrt{6}$)
for the large deviation of the minimum (maximum) eigenvalue. Thus, it is sufficient to consider the rescaled purities
$P_1$ and $P_3$ to find the desired relation between $\zeta_1$ and $\zeta_2$.
For given $\zeta_1$ the rescaled purity is $P_1=2-2\zeta_1+\zeta_1^2$. The parameter $\zeta_2$ for which the rescaled 
purity is $P_1$ one needs to solve for $P_1=P_3=(8\zeta_2-\zeta_2^2)/8$. Solving this quadratic equation one obtains
$\zeta_2=4\pm 2\sqrt{2(2-P_1)}=4\pm 2\sqrt{2\zeta_1-\zeta_1^2}$. Of these two solutions only 
$\zeta_2=4-2\sqrt{2(2\zeta_1-\zeta_1^2)}$ is valid while the other solution is invalid since it exceeds its 
upper limit which is four. 
For the special value of $\zeta_1=1/8$ the corresponding value of $\zeta_2$ for which the log negativity is same is 
approximately equal to $2.6307$. Using Eq.~(\ref{lognegform}) the log negativity is approximately equal to 
$0.0919$. These results can be observed in Fig.~\ref{fig:negativity}. It should be mentioned that these results are 
valid for the case $N_1\neq N_2$ since they depend only on $N=N_1N_2$ and $M$.

It is important to compare the results obtained in Secs.\ref{sec:BipartiteEntanglement} and 
\ref{sec:TripartiteEntanglement} using the Fig.~(\ref{fig:negativity}) and the bottom one in 
Fig.~(\ref{fig:RateFunction}). In can be seen that at $\zeta=1$ the von Neumann entropy is maximum while the log 
negativity is zero. As $\zeta$ goes away from $1$ the von Neumann entropy reduces while the log negativity 
increases outside the range $[1/2,4-\sqrt{6}]$. This behavior can be understood using the monogamous nature of the
entanglement \cite{Coffman}. It says that if two subsystems (here subsystems $1$ and $2$) have maximum quantum 
corrections then they (either $1$ or $2$) cannot be correlated at all with a third system (here subsystem $B$). 
This also implies that the joint system of $1$ and $2$ together also cannot be correlated at all with 
the third system. Monogamy of entanglement holds for each and every quantum state which implies it will also 
hold on an average. This is what is observed from these figures. 
It should be noted that this is a qualitative observation and a quantitative understanding demands thorough 
investigation.

\section{SUMMARY AND CONCLUSIONS}
 
This paper has studied the large deviations of the minimum Schmidt eigenvalue in a large bipartite system, denoted 
as $A$ and $B$. The state of the system is pure and chosen randomly from the uniform Haar measure. This eigenvalue 
play an important role in the study of entanglement between the two subsystems. Using the Coulomb gas method, the 
large deviation function for the minimum eigenvalue and the associated equilibrium charge density is derived. 
Our results hold for all the values of the Dyson index. These 
analytical expressions are found to agree very well with the Monte Carlo simulations. Thus, with this density the
generalization of the Marcenko-Pastur function is given when there are large deviations in the minimum Schmidt 
eigenvalue. In this paper the case of equal dimensions ($N=M$) of subsystems $A$ and $B$ is studied. 

The effect of the large deviations of both maximum and minimum eigenvalue is studied on the entanglement between 
$A$ and $B$ by using the von Neumann entropy. For this the equilibrium Coulomb charge density obtained for the large 
deviations of the minimum eigenvalue in this paper and the corresponding result for the maximum eigenvalue from 
earlier work in Ref.\cite{Nadal11} is used. In the case of large deviations of the maximum eigenvalue analytical expression 
for the entropy is 
derived using {\it Mathematica~9}, while the same for the minimum eigenvalue remains an open question. The entropy 
in the later case is obtained by numerical integration. These entropies are found to agree very well with the 
Monte Carlo 
simulations. The entropy corresponding to the large deviations of the maximum eigenvalue is continuous and once 
differentiable, but the second derivative is discontinuous at $\zeta=4/3$.
This is due to the transition in the density of states occurring at same $\zeta$ because of the 
large deviations in the maximum eigenvalue \cite{Nadal11}.

One of the subsystem is further divided into two parts, denoted as $1$ and $2$. The effect of the large deviations 
is also studied on the entanglement, measured using the log negativity, between $1$ and $2$. 
It is found that the state of the subsystem undergoes a NPT-PPT transition. 
The transition takes place at $\zeta=0.5$ ($\zeta=4-\sqrt{6}$) for the large deviations of the minimum 
(maximum) eigenvalue. To be precise, when $\zeta>1/2$ ($\zeta<4-\sqrt{6}$) for the large deviations of the 
minimum (maximum) eigenvalue the states are dominantly PPT, the critical barrier position being $\zeta=1/2$ 
($\zeta=4-\sqrt{6}$).  

It is found numerically that the density of states of the reduced density matrix of subsystems after PT is 
close to the Wigner semicircle law when there are large deviations in the extreme Schmidt eigenvalues. 
The skewness of the semicircle is minimum for the symmetric case $N_1=N_2$. Earlier work in Ref.\cite{Uday12} has 
shown the same when there are no such large deviations. Thus, our work shows the robustness of the Wigner 
semicircle law after PT even in the presence of large deviations in the extreme eigenvalues before PT. 
A simple random 
matrix model from the same work in Ref.\cite{Uday12} is used successfully to capture the NPT-PPT transition as 
well as the density of states after PT. One to one relationship between barrier positions $\zeta_1$ and 
$\zeta_2$, which corresponds to large deviations of minimum and maximum eigenvalues respectively, is found such 
that the entanglement between subsystems $1$ and $2$ is same for both the positions.
Results of bipartite and tripartite entanglement are interpreted qualitatively 
from the perspective of monogamous nature of the entanglement.

\section{Acknowledgments}
Author is very grateful to acknowledge many discussions with Arul Lakshminarayan and Karol \.Z{}yczkowski.
Author is happy to acknowledge many discussions with M. S. Santhanam and T. S. Mahesh. 
The author thanks C. S. Sudheer Kumar, G. Khairnar, H. Tekur, and S. Paul for carefully reading the manuscript. 
The author acknowledges the funding received from Department of Science and Technology, India under the scheme 
Science and Engineering Research Board (SERB) National Post Doctoral Fellowship (NPDF) file Number PDF/2015/00050.

\bibliography{reference22013,reference22}

\begin{thebibliography}{10}%
\makeatletter
\providecommand \@ifxundefined [1]{%
 \ifx #1\undefined \expandafter \@firstoftwo
 \else \expandafter \@secondoftwo
\fi
}%
\providecommand \@ifnum [1]{%
 \ifnum #1\expandafter \@firstoftwo
 \else \expandafter \@secondoftwo
\fi
}%
\providecommand \enquote [1]{``#1''}%
\providecommand \bibnamefont  [1]{#1}%
\providecommand \bibfnamefont [1]{#1}%
\providecommand \citenamefont [1]{#1}%
\providecommand\href[0]{\@sanitize\@href}%
\providecommand\@href[1]{\endgroup\@@startlink{#1}\endgroup\@@href}%
\providecommand\@@href[1]{#1\@@endlink}%
\providecommand \@sanitize [0]{\begingroup\catcode`\&12\catcode`\#12\relax}%
\@ifxundefined \pdfoutput {\@firstoftwo}{%
 \@ifnum{\z@=\pdfoutput}{\@firstoftwo}{\@secondoftwo}%
}{%
 \providecommand\@@startlink[1]{\leavevmode\special{html:<a href="#1">}}%
 \providecommand\@@endlink[0]{\special{html:</a>}}%
}{%
 \providecommand\@@startlink[1]{%
  \leavevmode
  \pdfstartlink
   attr{/Border[0 0 1 ]/H/I/C[0 1 1]}%
   user{/Subtype/Link/A<</Type/Action/S/URI/URI(#1)>>}%
  \relax
 }%
 \providecommand\@@endlink[0]{\pdfendlink}%
}%
\providecommand \url  [0]{\begingroup\@sanitize \@url }%
\providecommand \@url [1]{\endgroup\@href {#1}{\urlprefix}}%
\providecommand \urlprefix [0]{URL }%
\providecommand \Eprint[0]{\href }%
\@ifxundefined \urlstyle {%
  \providecommand \doi [1]{doi:\discretionary{}{}{}#1}%
}{%
  \providecommand \doi [0]{doi:\discretionary{}{}{}\begingroup
  \urlstyle{rm}\Url }%
}%
\providecommand \doibase [0]{http://dx.doi.org/}%
\providecommand \Doi[1]{\href{\doibase#1}}%
\providecommand \bibAnnote [3]{%
  \BibitemShut{#1}%
  \begin{quotation}\noindent
    \textsc{Key:}\ #2\\\textsc{Annotation:}\ #3%
  \end{quotation}%
}%
\providecommand \bibAnnoteFile [2]{%
  \IfFileExists{#2}{\bibAnnote {#1} {#2} {\input{#2}}}{}%
}%
\providecommand \typeout [0]{\immediate \write \m@ne }%
\providecommand \selectlanguage [0]{\@gobble}%
\providecommand \bibinfo [0]{\@secondoftwo}%
\providecommand \bibfield [0]{\@secondoftwo}%
\providecommand \translation [1]{[#1]}%
\providecommand \BibitemOpen[0]{}%
\providecommand \bibitemStop [0]{}%
\providecommand \bibitemNoStop [0]{.\EOS\space}%
\providecommand \EOS [0]{\spacefactor3000\relax}%
\providecommand \BibitemShut [1]{\csname bibitem#1\endcsname}%
\bibitem{Hollander}%
  \BibitemOpen
  \bibfield{author}{%
  \bibinfo {author} {\bibfnamefont{F.~D.}\ \bibnamefont{Hollander}},\ }%
  \emph{\bibinfo {title} {Large Deviations}}\ (\bibinfo {publisher} {American
  Mathematical Society},\ \bibinfo {year} {2000})%
  \bibAnnoteFile{NoStop}{Hollander}%
\bibitem{Vivo07}%
  \BibitemOpen
  \bibfield{author}{%
  \bibinfo {author} {\bibfnamefont{P.}~\bibnamefont{Vivo}}, \bibinfo {author}
  {\bibfnamefont{S.~N.}\ \bibnamefont{Majumdar}},\ and\ \bibinfo {author}
  {\bibfnamefont{O.}~\bibnamefont{Bohigas}},\ }%
  \bibfield{journal}{%
  \bibinfo {journal} {J. Phys. A: Math. Theor.}\ }%
  \textbf{\bibinfo {volume} {40}},\ \bibinfo {pages} {4317} (\bibinfo {year}
  {2007})%
  \bibAnnoteFile{NoStop}{Vivo07}%
\bibitem{Satyalarge}%
  \BibitemOpen
  \bibfield{author}{%
  \bibinfo {author} {\bibfnamefont{D.~S.}\ \bibnamefont{Dean}}\ and\ \bibinfo
  {author} {\bibfnamefont{S.~N.}\ \bibnamefont{Majumdar}},\ }%
  \bibfield{journal}{%
  \bibinfo {journal} {Phys. Rev. Lett.}\ }%
  \textbf{\bibinfo {volume} {97}},\ \bibinfo {pages} {160201} (\bibinfo {year}
  {2006})%
  \bibAnnoteFile{NoStop}{Satyalarge}%
\bibitem{Satyalarge1}%
  \BibitemOpen
  \bibfield{author}{%
  \bibinfo {author} {\bibfnamefont{D.~S.}\ \bibnamefont{Dean}}\ and\ \bibinfo
  {author} {\bibfnamefont{S.~N.}\ \bibnamefont{Majumdar}},\ }%
  \bibfield{journal}{%
  \bibinfo {journal} {Phys. Rev. E}\ }%
  \textbf{\bibinfo {volume} {77}},\ \bibinfo {pages} {041108} (\bibinfo {year}
  {2008})%
  \bibAnnoteFile{NoStop}{Satyalarge1}%
\bibitem{Castillo2016}%
  \BibitemOpen
  \bibfield{author}{%
  \bibinfo {author} {\bibfnamefont{F.~L.}\ \bibnamefont{Metz}}\ and\ \bibinfo
  {author} {\bibfnamefont{I.}~\bibnamefont{P\'erez~Castillo}},\ }%
  \bibfield{journal}{%
  \bibinfo {journal} {Phys. Rev. Lett.}\ }%
  \textbf{\bibinfo {volume} {117}},\ \bibinfo {pages} {104101} (\bibinfo {year}
  {2016})%
  \bibAnnoteFile{NoStop}{Castillo2016}%
\bibitem{Castillo2010}%
  \BibitemOpen
  \bibfield{author}{%
  \bibinfo {author} {\bibfnamefont{E.}~\bibnamefont{Katzav}}\ and\ \bibinfo
  {author} {\bibfnamefont{I.}~\bibnamefont{P\'erez~Castillo}},\ }%
  \bibfield{journal}{%
  \bibinfo {journal} {Phys. Rev. E}\ }%
  \textbf{\bibinfo {volume} {82}},\ \bibinfo {pages} {040104} (\bibinfo {year}
  {2010})%
  \bibAnnoteFile{NoStop}{Castillo2010}%
\bibitem{SatyaIndex2009}%
  \BibitemOpen
  \bibfield{author}{%
  \bibinfo {author} {\bibfnamefont{S.~N.}\ \bibnamefont{Majumdar}}, \bibinfo
  {author} {\bibfnamefont{C.}~\bibnamefont{Nadal}}, \bibinfo {author}
  {\bibfnamefont{A.}~\bibnamefont{Scardicchio}},\ and\ \bibinfo {author}
  {\bibfnamefont{P.}~\bibnamefont{Vivo}},\ }%
  \bibfield{journal}{%
  \bibinfo {journal} {Phys. Rev. Lett.}\ }%
  \textbf{\bibinfo {volume} {103}},\ \bibinfo {pages} {220603} (\bibinfo {year}
  {2009})%
  \bibAnnoteFile{NoStop}{SatyaIndex2009}%
\bibitem{Nadal10}%
  \BibitemOpen
  \bibfield{author}{%
  \bibinfo {author} {\bibfnamefont{C.}~\bibnamefont{Nadal}}, \bibinfo {author}
  {\bibfnamefont{S.~N.}\ \bibnamefont{Majumdar}},\ and\ \bibinfo {author}
  {\bibfnamefont{M.}~\bibnamefont{Vergassola}},\ }%
  \bibfield{journal}{%
  \bibinfo {journal} {Phys. Rev. Lett.}\ }%
  \textbf{\bibinfo {volume} {104}},\ \bibinfo {pages} {110501} (\bibinfo {year}
  {2010})%
  \bibAnnoteFile{NoStop}{Nadal10}%
\bibitem{Arul08}%
  \BibitemOpen
  \bibfield{author}{%
  \bibinfo {author} {\bibfnamefont{A.}~\bibnamefont{Lakshminarayan}}, \bibinfo
  {author} {\bibfnamefont{S.}~\bibnamefont{Tomsovic}}, \bibinfo {author}
  {\bibfnamefont{O.}~\bibnamefont{Bohigas}},\ and\ \bibinfo {author}
  {\bibfnamefont{S.~N.}\ \bibnamefont{Majumdar}},\ }%
  \bibfield{journal}{%
  \bibinfo {journal} {Phys. Rev. Lett.}\ }%
  \textbf{\bibinfo {volume} {100}},\ \bibinfo {pages} {044103} (\bibinfo {year}
  {2008})%
  \bibAnnoteFile{NoStop}{Arul08}%
\bibitem{majumdar08}%
  \BibitemOpen
  \bibfield{author}{%
  \bibinfo {author} {\bibfnamefont{S.~N.}\ \bibnamefont{Majumdar}}, \bibinfo
  {author} {\bibfnamefont{O.}~\bibnamefont{Bohigas}},\ and\ \bibinfo {author}
  {\bibfnamefont{A.}~\bibnamefont{Lakshminarayan}},\ }%
  \bibfield{journal}{%
  \bibinfo {journal} {J. Stat. Phys.}\ }%
  \textbf{\bibinfo {volume} {131}},\ \bibinfo {pages} {33} (\bibinfo {year}
  {2008})%
  \bibAnnoteFile{NoStop}{majumdar08}%
\bibitem{Nadal11}%
  \BibitemOpen
  \bibfield{author}{%
  \bibinfo {author} {\bibfnamefont{C.}~\bibnamefont{Nadal}}, \bibinfo {author}
  {\bibfnamefont{S.~N.}\ \bibnamefont{Majumdar}},\ and\ \bibinfo {author}
  {\bibfnamefont{M.}~\bibnamefont{Vergassola}},\ }%
  \bibfield{journal}{%
  \bibinfo {journal} {J. Stat. Phys.}\ }%
  \textbf{\bibinfo {volume} {142}},\ \bibinfo {pages} {403} (\bibinfo {year}
  {2011})%
  \bibAnnoteFile{NoStop}{Nadal11}%
\bibitem{Uday12}%
  \BibitemOpen
  \bibfield{author}{%
  \bibinfo {author} {\bibfnamefont{U.~T.}\ \bibnamefont{Bhosale}}, \bibinfo
  {author} {\bibfnamefont{S.}~\bibnamefont{Tomsovic}},\ and\ \bibinfo {author}
  {\bibfnamefont{A.}~\bibnamefont{Lakshminarayan}},\ }%
  \bibfield{journal}{%
  \bibinfo {journal} {Phys. Rev. A}\ }%
  \textbf{\bibinfo {volume} {85}},\ \bibinfo {pages} {062331} (\bibinfo {year}
  {2012})%
  \bibAnnoteFile{NoStop}{Uday12}%
\bibitem{Karol2017}%
  \BibitemOpen
  \bibfield{author}{%
  \bibinfo {author} {\bibfnamefont{K.}~\bibnamefont{Szyma\'{n}ski}}, \bibinfo
  {author} {\bibfnamefont{B.}~\bibnamefont{Collins}}, \bibinfo {author}
  {\bibfnamefont{T.}~\bibnamefont{Szarek}},\ and\ \bibinfo {author}
  {\bibfnamefont{K.}~\bibnamefont{\ifmmode~\dot{Z}\else \.{Z}\fi{}yczkowski}},\
  }%
  \bibfield{journal}{%
  \bibinfo {journal} {Journal of Physics A: Mathematical and Theoretical}\ }%
  \textbf{\bibinfo {volume} {50}},\ \bibinfo {pages} {255206} (\bibinfo {year}
  {2017})%
  \bibAnnoteFile{NoStop}{Karol2017}%
\bibitem{Vivo2011}%
  \BibitemOpen
  \bibfield{author}{%
  \bibinfo {author} {\bibfnamefont{P.}~\bibnamefont{Vivo}},\ }%
  \bibfield{journal}{%
  \bibinfo {journal} {J. Phys. A: Math. Theor.},\ \bibinfo {pages} {P01022}}%
   (\bibinfo {year} {2011})%
  \bibAnnoteFile{NoStop}{Vivo2011}%
\bibitem{Chavez2015}%
  \BibitemOpen
  \bibfield{author}{%
  \bibinfo {author} {\bibfnamefont{M.}~\bibnamefont{Chavez}}, \bibinfo {author}
  {\bibfnamefont{M.}~\bibnamefont{Ghil}},\ and\ \bibinfo {author}
  {\bibfnamefont{J.}~\bibnamefont{Urrutia-Fucugauchi}},\ }%
  \emph{\bibinfo {title} {Extreme Events: Observations, Modeling, and
  Economics}}\ (\bibinfo {publisher} {Wiley},\ \bibinfo {year} {2015})%
  \bibAnnoteFile{NoStop}{Chavez2015}%
\bibitem{Sergio2006}%
  \BibitemOpen
  \bibfield{author}{%
  \bibinfo {author} {\bibfnamefont{S.}~\bibnamefont{Albeverio}}, \bibinfo
  {author} {\bibfnamefont{V.}~\bibnamefont{Jentsch}},\ and\ \bibinfo {author}
  {\bibfnamefont{H.}~\bibnamefont{Kantz}},\ }%
  \emph{\bibinfo {title} {Extreme Events in Nature and Society}}\ (\bibinfo
  {publisher} {Springer-Verlag Berlin Heidelberg},\ \bibinfo {year} {2006})%
  \bibAnnoteFile{NoStop}{Sergio2006}%
\bibitem{wilksbook}%
  \BibitemOpen
  \bibfield{author}{%
  \bibinfo {author} {\bibfnamefont{S.~S.}\ \bibnamefont{Wilks}},\ }%
  \emph{\bibinfo {title} {Mathematical Statistics}}\ (\bibinfo {publisher}
  {Princeton University Press, New Jersey},\ \bibinfo {year} {1947})%
  \bibAnnoteFile{NoStop}{wilksbook}%
\bibitem{Fukunaga}%
  \BibitemOpen
  \bibfield{author}{%
  \bibinfo {author} {\bibfnamefont{K.}~\bibnamefont{Fukunaga}},\ }%
  \emph{\bibinfo {title} {Introduction to Statistical Pattern Recognition}}\
  (\bibinfo {publisher} {Elsevier, New York},\ \bibinfo {year} {1990})%
  \bibAnnoteFile{NoStop}{Fukunaga}%
\bibitem{Fridman12}%
  \BibitemOpen
  \bibfield{author}{%
  \bibinfo {author} {\bibfnamefont{M.}~\bibnamefont{Fridman}}, \bibinfo
  {author} {\bibfnamefont{R.}~\bibnamefont{Pugatch}}, \bibinfo {author}
  {\bibfnamefont{M.}~\bibnamefont{Nixon}}, \bibinfo {author}
  {\bibfnamefont{A.~A.}\ \bibnamefont{Friesem}},\ and\ \bibinfo {author}
  {\bibfnamefont{N.}~\bibnamefont{Davidson}},\ }%
  \bibfield{journal}{%
  \bibinfo {journal} {Phys. Rev. E}\ }%
  \textbf{\bibinfo {volume} {85}},\ \bibinfo {pages} {020101} (\bibinfo {year}
  {2012})%
  \bibAnnoteFile{NoStop}{Fridman12}%
\bibitem{Jonathan2017}%
  \BibitemOpen
  \bibfield{author}{%
  \bibinfo {author} {\bibfnamefont{J.~G.}\ \bibnamefont{Richens}}, \bibinfo
  {author} {\bibfnamefont{J.~H.}\ \bibnamefont{Selby}},\ and\ \bibinfo {author}
  {\bibfnamefont{S.~W.}\ \bibnamefont{Al-Safi}},\ }%
  \bibfield{journal}{%
  \bibinfo {journal} {Phys. Rev. Lett.}\ }%
  \textbf{\bibinfo {volume} {119}},\ \bibinfo {pages} {080503} (\bibinfo {year}
  {2017})%
  \bibAnnoteFile{NoStop}{Jonathan2017}%
\bibitem{Horodeckirpm}%
  \BibitemOpen
  \bibfield{author}{%
  \bibinfo {author} {\bibfnamefont{R.}~\bibnamefont{Horodecki}}, \bibinfo
  {author} {\bibfnamefont{P.}~\bibnamefont{Horodecki}}, \bibinfo {author}
  {\bibfnamefont{M.}~\bibnamefont{Horodecki}},\ and\ \bibinfo {author}
  {\bibfnamefont{K.}~\bibnamefont{Horodecki}},\ }%
  \bibfield{journal}{%
  \bibinfo {journal} {Rev. Mod. Phys.}\ }%
  \textbf{\bibinfo {volume} {81}},\ \bibinfo {pages} {865} (\bibinfo {year}
  {2009})%
  \bibAnnoteFile{NoStop}{Horodeckirpm}%
\bibitem{bennett93}%
  \BibitemOpen
  \bibfield{author}{%
  \bibinfo {author} {\bibfnamefont{C.~H.}\ \bibnamefont{Bennett}}, \bibinfo
  {author} {\bibfnamefont{G.}~\bibnamefont{Brassard}}, \bibinfo {author}
  {\bibfnamefont{C.}~\bibnamefont{Cr\'epeau}}, \bibinfo {author}
  {\bibfnamefont{R.}~\bibnamefont{Jozsa}}, \bibinfo {author}
  {\bibfnamefont{A.}~\bibnamefont{Peres}},\ and\ \bibinfo {author}
  {\bibfnamefont{W.~K.}\ \bibnamefont{Wootters}},\ }%
  \bibfield{journal}{%
  \bibinfo {journal} {Phys. Rev. Lett.}\ }%
  \textbf{\bibinfo {volume} {70}},\ \bibinfo {pages} {1895} (\bibinfo {year}
  {1993})%
  \bibAnnoteFile{NoStop}{bennett93}%
\bibitem{Superdense}%
  \BibitemOpen
  \bibfield{author}{%
  \bibinfo {author} {\bibfnamefont{C.~H.}\ \bibnamefont{Bennett}}\ and\
  \bibinfo {author} {\bibfnamefont{S.~J.}\ \bibnamefont{Wiesner}},\ }%
  \bibfield{journal}{%
  \bibinfo {journal} {Phys. Rev. Lett.}\ }%
  \textbf{\bibinfo {volume} {69}},\ \bibinfo {pages} {2881} (\bibinfo {year}
  {1992})%
  \bibAnnoteFile{NoStop}{Superdense}%
\bibitem{Bennett96}%
  \BibitemOpen
  \bibfield{author}{%
  \bibinfo {author} {\bibfnamefont{C.~H.}\ \bibnamefont{Bennett}}, \bibinfo
  {author} {\bibfnamefont{H.~J.}\ \bibnamefont{Bernstein}}, \bibinfo {author}
  {\bibfnamefont{S.}~\bibnamefont{Popescu}},\ and\ \bibinfo {author}
  {\bibfnamefont{B.}~\bibnamefont{Schumacher}},\ }%
  \bibfield{journal}{%
  \bibinfo {journal} {Phys. Rev. A}\ }%
  \textbf{\bibinfo {volume} {53}},\ \bibinfo {pages} {2046} (\bibinfo {year}
  {1996})%
  \bibAnnoteFile{NoStop}{Bennett96}%
\bibitem{Zyczkowski06}%
  \BibitemOpen
  \bibfield{author}{%
  \bibinfo {author} {\bibfnamefont{I.}~\bibnamefont{Bengtsson}}\ and\ \bibinfo
  {author} {\bibfnamefont{K.}~\bibnamefont{Zyczkowski}},\ }%
  \emph{\bibinfo {title} {Geometry of Quantum States: An Introduction to
  Quantum Entanglement}}\ (\bibinfo {publisher} {Cambridge University Press,
  Cambridge},\ \bibinfo {year} {2006})%
  \bibAnnoteFile{NoStop}{Zyczkowski06}%
\bibitem{Demmel1988}%
  \BibitemOpen
  \bibfield{author}{%
  \bibinfo {author} {\bibfnamefont{J.~W.}\ \bibnamefont{Demmel}},\ }%
  \bibfield{journal}{%
  \bibinfo {journal} {Math. Comput.}\ }%
  \textbf{\bibinfo {volume} {50}},\ \bibinfo {pages} {449} (\bibinfo {year}
  {1988})%
  \bibAnnoteFile{NoStop}{Demmel1988}%
\bibitem{Edelman1992}%
  \BibitemOpen
  \bibfield{author}{%
  \bibinfo {author} {\bibfnamefont{A.}~\bibnamefont{Edelman}},\ }%
  \bibfield{journal}{%
  \bibinfo {journal} {Math. Comput.}\ }%
  \textbf{\bibinfo {volume} {58}},\ \bibinfo {pages} {185} (\bibinfo {year}
  {1992})%
  \bibAnnoteFile{NoStop}{Edelman1992}%
\bibitem{MarkoZ07}%
  \BibitemOpen
  \bibfield{author}{%
  \bibinfo {author} {\bibfnamefont{M.}~\bibnamefont{Znidaric}},\ }%
  \bibfield{journal}{%
  \bibinfo {journal} {J. Phys. A: Math. Theor.}\ }%
  \textbf{\bibinfo {volume} {40}},\ \bibinfo {pages} {F105} (\bibinfo {year}
  {2007})%
  \bibAnnoteFile{NoStop}{MarkoZ07}%
\bibitem{YangChen2010}%
  \BibitemOpen
  \bibfield{author}{%
  \bibinfo {author} {\bibfnamefont{Y.}~\bibnamefont{Chen}}, \bibinfo {author}
  {\bibfnamefont{D.-Z.}\ \bibnamefont{Liu}},\ and\ \bibinfo {author}
  {\bibfnamefont{D.-S.}\ \bibnamefont{Zhou}},\ }%
  \bibfield{journal}{%
  \bibinfo {journal} {J. Phys. A: Math. Theor.}\ }%
  \textbf{\bibinfo {volume} {43}},\ \bibinfo {pages} {315303} (\bibinfo {year}
  {2010})%
  \bibAnnoteFile{NoStop}{YangChen2010}%
\bibitem{Nadler2011}%
  \BibitemOpen
  \bibfield{author}{%
  \bibinfo {author} {\bibfnamefont{B.}~\bibnamefont{Nadler}},\ }%
  \bibfield{journal}{%
  \bibinfo {journal} {J. Multivariate Anal.}\ }%
  \textbf{\bibinfo {volume} {102}},\ \bibinfo {pages} {363} (\bibinfo {year}
  {2011})%
  \bibAnnoteFile{NoStop}{Nadler2011}%
\bibitem{Marcenko}%
  \BibitemOpen
  \bibfield{author}{%
  \bibinfo {author} {\bibfnamefont{V.~A.}\ \bibnamefont{Marcenko}}\ and\
  \bibinfo {author} {\bibfnamefont{L.~A.}\ \bibnamefont{Pastur}},\ }%
  \bibfield{journal}{%
  \bibinfo {journal} {Math. USSR-Sb}\ }%
  \textbf{\bibinfo {volume} {1}},\ \bibinfo {pages} {457} (\bibinfo {year}
  {1967})%
  \bibAnnoteFile{NoStop}{Marcenko}%
\bibitem{Borot2012}%
  \BibitemOpen
  \bibfield{author}{%
  \bibinfo {author} {\bibfnamefont{G.}~\bibnamefont{Borot}}\ and\ \bibinfo
  {author} {\bibfnamefont{C.}~\bibnamefont{Nadal}},\ }%
  \bibfield{journal}{%
  \bibinfo {journal} {J. Phys. A: Math. Theor.}\ }%
  \textbf{\bibinfo {volume} {45}},\ \bibinfo {pages} {075209} (\bibinfo {year}
  {1997})%
  \bibAnnoteFile{NoStop}{Borot2012}%
\bibitem{Bouchaud}%
  \BibitemOpen
  \bibfield{author}{%
  \bibinfo {author} {\bibfnamefont{B.}~\bibnamefont{J-P}}\ and\ \bibinfo
  {author} {\bibfnamefont{P.}~\bibnamefont{M}},\ }%
  \emph{\bibinfo {title} {Theory of Financial Risks}}\ (\bibinfo {publisher}
  {Cambridge: Cambridge University Press},\ \bibinfo {year} {2001})%
  \bibAnnoteFile{NoStop}{Bouchaud}%
\bibitem{Fyodorov1997}%
  \BibitemOpen
  \bibfield{author}{%
  \bibinfo {author} {\bibfnamefont{Y.~V.}\ \bibnamefont{Fyodorov}}\ and\
  \bibinfo {author} {\bibfnamefont{H.-J.}\ \bibnamefont{Sommers}},\ }%
  \bibfield{journal}{%
  \bibinfo {journal} {J. Math. Phys.}\ }%
  \textbf{\bibinfo {volume} {38}},\ \bibinfo {pages} {1918} (\bibinfo {year}
  {1997})%
  \bibAnnoteFile{NoStop}{Fyodorov1997}%
\bibitem{Fyodorov1999}%
  \BibitemOpen
  \bibfield{author}{%
  \bibinfo {author} {\bibfnamefont{Y.~V.}\ \bibnamefont{Fyodorov}}\ and\
  \bibinfo {author} {\bibfnamefont{B.~A.}\ \bibnamefont{Khoruzhenko}},\ }%
  \bibfield{journal}{%
  \bibinfo {journal} {Phys. Rev. Lett.}\ }%
  \textbf{\bibinfo {volume} {83}},\ \bibinfo {pages} {65} (\bibinfo {year}
  {1999})%
  \bibAnnoteFile{NoStop}{Fyodorov1999}%
\bibitem{Shuryak93}%
  \BibitemOpen
  \bibfield{author}{%
  \bibinfo {author} {\bibfnamefont{E.}~\bibnamefont{Shuryak}}\ and\ \bibinfo
  {author} {\bibfnamefont{J.}~\bibnamefont{Verbaarschot}},\ }%
  \bibfield{journal}{%
  \bibinfo {journal} {Nucl. Phys. A}\ }%
  \textbf{\bibinfo {volume} {560}},\ \bibinfo {pages} {306} (\bibinfo {year}
  {1993})%
  \bibAnnoteFile{NoStop}{Shuryak93}%
\bibitem{Verbaarschot94}%
  \BibitemOpen
  \bibfield{author}{%
  \bibinfo {author} {\bibfnamefont{J.}~\bibnamefont{Verbaarschot}},\ }%
  \bibfield{journal}{%
  \bibinfo {journal} {Phys. Rev. Lett.}\ }%
  \textbf{\bibinfo {volume} {72}},\ \bibinfo {pages} {2531} (\bibinfo {year}
  {1994})%
  \bibAnnoteFile{NoStop}{Verbaarschot94}%
\bibitem{Maslov2001}%
  \BibitemOpen
  \bibfield{author}{%
  \bibinfo {author} {\bibfnamefont{S.}~\bibnamefont{Maslov}}\ and\ \bibinfo
  {author} {\bibfnamefont{Y.-C.}\ \bibnamefont{Zhang}},\ }%
  \bibfield{journal}{%
  \bibinfo {journal} {Phys. Rev. Lett.}\ }%
  \textbf{\bibinfo {volume} {87}},\ \bibinfo {pages} {248701} (\bibinfo {year}
  {2001})%
  \bibAnnoteFile{NoStop}{Maslov2001}%
\bibitem{Tracy1}%
  \BibitemOpen
  \bibfield{author}{%
  \bibinfo {author} {\bibfnamefont{C.}~\bibnamefont{Tracy}}\ and\ \bibinfo
  {author} {\bibfnamefont{H.}~\bibnamefont{Widom}},\ }%
  \bibfield{journal}{%
  \bibinfo {journal} {Commun. Math. Phys.}\ }%
  \textbf{\bibinfo {volume} {159}},\ \bibinfo {pages} {151} (\bibinfo {year}
  {1994})%
  \bibAnnoteFile{NoStop}{Tracy1}%
\bibitem{Tracy2}%
  \BibitemOpen
  \bibfield{author}{%
  \bibinfo {author} {\bibfnamefont{C.}~\bibnamefont{Tracy}}\ and\ \bibinfo
  {author} {\bibfnamefont{H.}~\bibnamefont{Widom}},\ }%
  \bibfield{journal}{%
  \bibinfo {journal} {Commun. Math. Phys.}\ }%
  \textbf{\bibinfo {volume} {177}},\ \bibinfo {pages} {727} (\bibinfo {year}
  {1996})%
  \bibAnnoteFile{NoStop}{Tracy2}%
\bibitem{Satyalargeright}%
  \BibitemOpen
  \bibfield{author}{%
  \bibinfo {author} {\bibfnamefont{S.~N.}\ \bibnamefont{Majumdar}}\ and\
  \bibinfo {author} {\bibfnamefont{M.}~\bibnamefont{Vergassola}},\ }%
  \bibfield{journal}{%
  \bibinfo {journal} {Phys. Rev. Lett.}\ }%
  \textbf{\bibinfo {volume} {102}},\ \bibinfo {pages} {060601} (\bibinfo {year}
  {2009})%
  \bibAnnoteFile{NoStop}{Satyalargeright}%
\bibitem{Dyson621}%
  \BibitemOpen
  \bibfield{author}{%
  \bibinfo {author} {\bibfnamefont{F.~J.}\ \bibnamefont{Dyson}},\ }%
  \bibfield{journal}{%
  \bibinfo {journal} {J. Math. Phys.}\ }%
  \textbf{\bibinfo {volume} {3}},\ \bibinfo {pages} {140} (\bibinfo {year}
  {1962})%
  \bibAnnoteFile{NoStop}{Dyson621}%
\bibitem{Dyson622}%
  \BibitemOpen
  \bibfield{author}{%
  \bibinfo {author} {\bibfnamefont{F.~J.}\ \bibnamefont{Dyson}},\ }%
  \bibfield{journal}{%
  \bibinfo {journal} {J. Math. Phys.}\ }%
  \textbf{\bibinfo {volume} {3}},\ \bibinfo {pages} {157} (\bibinfo {year}
  {1962})%
  \bibAnnoteFile{NoStop}{Dyson622}%
\bibitem{Dyson623}%
  \BibitemOpen
  \bibfield{author}{%
  \bibinfo {author} {\bibfnamefont{F.~J.}\ \bibnamefont{Dyson}},\ }%
  \bibfield{journal}{%
  \bibinfo {journal} {J. Math. Phys.}\ }%
  \textbf{\bibinfo {volume} {3}},\ \bibinfo {pages} {166} (\bibinfo {year}
  {1962})%
  \bibAnnoteFile{NoStop}{Dyson623}%
\bibitem{Nadal11Tracy}%
  \BibitemOpen
  \bibfield{author}{%
  \bibinfo {author} {\bibfnamefont{C.}~\bibnamefont{Nadal}}\ and\ \bibinfo
  {author} {\bibfnamefont{S.~N.}\ \bibnamefont{Majumdar}},\ }%
  \bibfield{journal}{%
  \bibinfo {journal} {J. Stat. Mech.}\ }%
  \textbf{\bibinfo {volume} {2011}},\ \bibinfo {pages} {P04001} (\bibinfo
  {year} {2011})%
  \bibAnnoteFile{NoStop}{Nadal11Tracy}%
\bibitem{Cunden2016}%
  \BibitemOpen
  \bibfield{author}{%
  \bibinfo {author} {\bibfnamefont{F.~D.}\ \bibnamefont{Cunden}}, \bibinfo
  {author} {\bibfnamefont{P.}~\bibnamefont{Facchi}},\ and\ \bibinfo {author}
  {\bibfnamefont{P.}~\bibnamefont{Vivo}},\ }%
  \bibfield{journal}{%
  \bibinfo {journal} {J. Phys. A: Math. Theor.}\ }%
  \textbf{\bibinfo {volume} {49}},\ \bibinfo {pages} {135202} (\bibinfo {year}
  {2016})%
  \bibAnnoteFile{NoStop}{Cunden2016}%
\bibitem{Marino2014}%
  \BibitemOpen
  \bibfield{author}{%
  \bibinfo {author} {\bibfnamefont{R.}~\bibnamefont{Marino}}, \bibinfo {author}
  {\bibfnamefont{S.~N.}\ \bibnamefont{Majumdar}}, \bibinfo {author}
  {\bibfnamefont{G.}~\bibnamefont{Schehr}},\ and\ \bibinfo {author}
  {\bibfnamefont{P.}~\bibnamefont{Vivo}},\ }%
  \bibfield{journal}{%
  \bibinfo {journal} {J. Phys. A: Math. Theor.}\ }%
  \textbf{\bibinfo {volume} {47}},\ \bibinfo {pages} {055001} (\bibinfo {year}
  {2014})%
  \bibAnnoteFile{NoStop}{Marino2014}%
\bibitem{llyod}%
  \BibitemOpen
  \bibfield{author}{%
  \bibinfo {author} {\bibfnamefont{S.}~\bibnamefont{Lloyd}}\ and\ \bibinfo
  {author} {\bibfnamefont{H.}~\bibnamefont{Pagels}},\ }%
  \bibfield{journal}{%
  \bibinfo {journal} {Ann. Phys.}\ }%
  \textbf{\bibinfo {volume} {188}},\ \bibinfo {pages} {186} (\bibinfo {year}
  {1988})%
  \bibAnnoteFile{NoStop}{llyod}%
\bibitem{Sommers}%
  \BibitemOpen
  \bibfield{author}{%
  \bibinfo {author} {\bibfnamefont{K.}~\bibnamefont{Zyczkowski}}\ and\ \bibinfo
  {author} {\bibfnamefont{H.-J.}\ \bibnamefont{Sommers}},\ }%
  \bibfield{journal}{%
  \bibinfo {journal} {J. Phys. A: Math. Gen.}\ }%
  \textbf{\bibinfo {volume} {34}},\ \bibinfo {pages} {7111} (\bibinfo {year}
  {2001})%
  \bibAnnoteFile{NoStop}{Sommers}%
\bibitem{Sommers04}%
  \BibitemOpen
  \bibfield{author}{%
  \bibinfo {author} {\bibfnamefont{H.-J.}\ \bibnamefont{Sommers}}\ and\
  \bibinfo {author} {\bibfnamefont{K.}~\bibnamefont{Zyczkowski}},\ }%
  \bibfield{journal}{%
  \bibinfo {journal} {J. Phys. A: Math. Gen.}\ }%
  \textbf{\bibinfo {volume} {37}},\ \bibinfo {pages} {8457} (\bibinfo {year}
  {2004})%
  \bibAnnoteFile{NoStop}{Sommers04}%
\bibitem{Lubkin}%
  \BibitemOpen
  \bibfield{author}{%
  \bibinfo {author} {\bibfnamefont{E.}~\bibnamefont{Lubkin}},\ }%
  \bibfield{journal}{%
  \bibinfo {journal} {J. Math. Phys.}\ }%
  \textbf{\bibinfo {volume} {19}},\ \bibinfo {pages} {1028} (\bibinfo {year}
  {1978})%
  \bibAnnoteFile{NoStop}{Lubkin}%
\bibitem{Page}%
  \BibitemOpen
  \bibfield{author}{%
  \bibinfo {author} {\bibfnamefont{D.}~\bibnamefont{Page}},\ }%
  \bibfield{journal}{%
  \bibinfo {journal} {Phys. Rev. Lett.}\ }%
  \textbf{\bibinfo {volume} {71}},\ \bibinfo {pages} {9} (\bibinfo {year}
  {1993})%
  \bibAnnoteFile{NoStop}{Page}%
\bibitem{Sen}%
  \BibitemOpen
  \bibfield{author}{%
  \bibinfo {author} {\bibfnamefont{S.}~\bibnamefont{Sen}},\ }%
  \bibfield{journal}{%
  \bibinfo {journal} {Phys. Rev. Lett.}\ }%
  \textbf{\bibinfo {volume} {77}},\ \bibinfo {pages} {1} (\bibinfo {year}
  {1996})%
  \bibAnnoteFile{NoStop}{Sen}%
\bibitem{Jorge}%
  \BibitemOpen
  \bibfield{author}{%
  \bibinfo {author} {\bibfnamefont{J.}~\bibnamefont{Sanchez-Ruiz}},\ }%
  \bibfield{journal}{%
  \bibinfo {journal} {Phys. Rev. E}\ }%
  \textbf{\bibinfo {volume} {52}},\ \bibinfo {pages} {5653} (\bibinfo {year}
  {1995})%
  \bibAnnoteFile{NoStop}{Jorge}%
\bibitem{Forresterbook}%
  \BibitemOpen
  \bibfield{author}{%
  \bibinfo {author} {\bibfnamefont{P.~J.}\ \bibnamefont{Forrester}},\ }%
  \emph{\bibinfo {title} {Log-Gases and Random Matrices}}\ (\bibinfo
  {publisher} {Princeton University Press, Princeton and Oxford},\ \bibinfo
  {year} {2010})%
  \bibAnnoteFile{NoStop}{Forresterbook}%
\bibitem{TricomiBook}%
  \BibitemOpen
  \bibfield{author}{%
  \bibinfo {author} {\bibfnamefont{F.~G.}\ \bibnamefont{Tricomi}},\ }%
  \emph{\bibinfo {title} {Integral Equations. Pure Appl. Math., vol. V.}}\
  (\bibinfo {publisher} {Interscience, London},\ \bibinfo {year} {1957})%
  \bibAnnoteFile{NoStop}{TricomiBook}%
\bibitem{MajumdarNumber2012}%
  \BibitemOpen
  \bibfield{author}{%
  \bibinfo {author} {\bibfnamefont{S.~N.}\ \bibnamefont{Majumdar}}\ and\
  \bibinfo {author} {\bibfnamefont{P.}~\bibnamefont{Vivo}},\ }%
  \bibfield{journal}{%
  \bibinfo {journal} {Phys. Rev. Lett.}\ }%
  \textbf{\bibinfo {volume} {108}},\ \bibinfo {pages} {200601} (\bibinfo {year}
  {2012})%
  \bibAnnoteFile{NoStop}{MajumdarNumber2012}%
\bibitem{SatyaMajumdar2011}%
  \BibitemOpen
  \bibfield{author}{%
  \bibinfo {author} {\bibfnamefont{S.~N.}\ \bibnamefont{Majumdar}}, \bibinfo
  {author} {\bibfnamefont{C.}~\bibnamefont{Nadal}}, \bibinfo {author}
  {\bibfnamefont{A.}~\bibnamefont{Scardicchio}},\ and\ \bibinfo {author}
  {\bibfnamefont{P.}~\bibnamefont{Vivo}},\ }%
  \bibfield{journal}{%
  \bibinfo {journal} {Phys. Rev. E}\ }%
  \textbf{\bibinfo {volume} {83}},\ \bibinfo {pages} {041105} (\bibinfo {year}
  {2011})%
  \bibAnnoteFile{NoStop}{SatyaMajumdar2011}%
\bibitem{Pierpaolo2008}%
  \BibitemOpen
  \bibfield{author}{%
  \bibinfo {author} {\bibfnamefont{P.}~\bibnamefont{Vivo}}, \bibinfo {author}
  {\bibfnamefont{S.~N.}\ \bibnamefont{Majumdar}},\ and\ \bibinfo {author}
  {\bibfnamefont{O.}~\bibnamefont{Bohigas}},\ }%
  \bibfield{journal}{%
  \bibinfo {journal} {Phys. Rev. Lett.}\ }%
  \textbf{\bibinfo {volume} {101}},\ \bibinfo {pages} {216809} (\bibinfo {year}
  {2008})%
  \bibAnnoteFile{NoStop}{Pierpaolo2008}%
\bibitem{Damle2011}%
  \BibitemOpen
  \bibfield{author}{%
  \bibinfo {author} {\bibfnamefont{K.}~\bibnamefont{Damle}}, \bibinfo {author}
  {\bibfnamefont{S.~N.}\ \bibnamefont{Majumdar}}, \bibinfo {author}
  {\bibfnamefont{V.}~\bibnamefont{Tripathi}},\ and\ \bibinfo {author}
  {\bibfnamefont{P.}~\bibnamefont{Vivo}},\ }%
  \bibfield{journal}{%
  \bibinfo {journal} {Phys. Rev. Lett.}\ }%
  \textbf{\bibinfo {volume} {107}},\ \bibinfo {pages} {177206} (\bibinfo {year}
  {2011})%
  \bibAnnoteFile{NoStop}{Damle2011}%
\bibitem{Grabsch2017}%
  \BibitemOpen
  \bibfield{author}{%
  \bibinfo {author} {\bibfnamefont{A.}~\bibnamefont{Grabsch}}, \bibinfo
  {author} {\bibfnamefont{S.~N.}\ \bibnamefont{Majumdar}},\ and\ \bibinfo
  {author} {\bibfnamefont{C.}~\bibnamefont{Texier}},\ }%
  \bibfield{journal}{%
  \bibinfo {journal} {J Stat Phys}\ }%
  \textbf{\bibinfo {volume} {167}},\ \bibinfo {pages} {234} (\bibinfo {year}
  {2017})%
  \bibAnnoteFile{NoStop}{Grabsch2017}%
\bibitem{SNMajumdar2014}%
  \BibitemOpen
  \bibfield{author}{%
  \bibinfo {author} {\bibfnamefont{S.~N.}\ \bibnamefont{Majumdar}}\ and\
  \bibinfo {author} {\bibfnamefont{G.}~\bibnamefont{Schehr}},\ }%
  \bibfield{journal}{%
  \bibinfo {journal} {J. Stat. Mech.}\ }%
  \textbf{\bibinfo {volume} {2014}},\ \bibinfo {pages} {P01012} (\bibinfo
  {year} {2014})%
  \bibAnnoteFile{NoStop}{SNMajumdar2014}%
\bibitem{Texier2013}%
  \BibitemOpen
  \bibfield{author}{%
  \bibinfo {author} {\bibfnamefont{C.}~\bibnamefont{Texier}}\ and\ \bibinfo
  {author} {\bibfnamefont{S.~N.}\ \bibnamefont{Majumdar}},\ }%
  \bibfield{journal}{%
  \bibinfo {journal} {Phys. Rev. Lett.}\ }%
  \textbf{\bibinfo {volume} {110}},\ \bibinfo {pages} {250602} (\bibinfo {year}
  {2013})%
  \bibAnnoteFile{NoStop}{Texier2013}%
\bibitem{Grabsch2016}%
  \BibitemOpen
  \bibfield{author}{%
  \bibinfo {author} {\bibfnamefont{A.}~\bibnamefont{Grabsch}}\ and\ \bibinfo
  {author} {\bibfnamefont{C.}~\bibnamefont{Texier}},\ }%
  \bibfield{journal}{%
  \bibinfo {journal} {J. Phys. A: Math. Theor.}\ }%
  \textbf{\bibinfo {volume} {49}},\ \bibinfo {pages} {465002} (\bibinfo {year}
  {2016})%
  \bibAnnoteFile{NoStop}{Grabsch2016}%
\bibitem{PierpaoloVivo2010}%
  \BibitemOpen
  \bibfield{author}{%
  \bibinfo {author} {\bibfnamefont{P.}~\bibnamefont{Vivo}}, \bibinfo {author}
  {\bibfnamefont{S.~N.}\ \bibnamefont{Majumdar}},\ and\ \bibinfo {author}
  {\bibfnamefont{O.}~\bibnamefont{Bohigas}},\ }%
  \bibfield{journal}{%
  \bibinfo {journal} {Phys. Rev. B}\ }%
  \textbf{\bibinfo {volume} {81}},\ \bibinfo {pages} {104202} (\bibinfo {year}
  {2010})%
  \bibAnnoteFile{NoStop}{PierpaoloVivo2010}%
\bibitem{Alan2007}%
  \BibitemOpen
  \bibfield{author}{%
  \bibinfo {author} {\bibfnamefont{A.~J.}\ \bibnamefont{Bray}}\ and\ \bibinfo
  {author} {\bibfnamefont{D.~S.}\ \bibnamefont{Dean}},\ }%
  \bibfield{journal}{%
  \bibinfo {journal} {Phys. Rev. Lett.}\ }%
  \textbf{\bibinfo {volume} {98}},\ \bibinfo {pages} {150201} (\bibinfo {year}
  {2007})%
  \bibAnnoteFile{NoStop}{Alan2007}%
\bibitem{RMTBook}%
  \BibitemOpen
  \bibfield{author}{%
  \bibinfo {author} {\bibfnamefont{S.~N.}\ \bibnamefont{Majumdar}},\ }%
  \emph{\bibinfo {title} {Extreme eigenvalues of Wishart matrices: application
  to entangled bipartite system}}\ (\bibinfo {publisher} {Akemann, G., Baik,
  J., Di Francesco P. (eds.) Handbook of Random Matrix Theory. Oxford
  University Press, London},\ \bibinfo {year} {2010})%
  \bibAnnoteFile{NoStop}{RMTBook}%
\bibitem{Aubrun12}%
  \BibitemOpen
  \bibfield{author}{%
  \bibinfo {author} {\bibfnamefont{G.}~\bibnamefont{Aubrun}}, \bibinfo {author}
  {\bibfnamefont{S.~J.}\ \bibnamefont{Szarek}},\ and\ \bibinfo {author}
  {\bibfnamefont{D.}~\bibnamefont{Ye}},\ }%
  \bibfield{journal}{%
  \bibinfo {journal} {Phys. Rev. A}\ }%
  \textbf{\bibinfo {volume} {85}},\ \bibinfo {pages} {030302(R)} (\bibinfo
  {year} {2012})%
  \bibAnnoteFile{NoStop}{Aubrun12}%
\bibitem{Aubrun2014}%
  \BibitemOpen
  \bibfield{author}{%
  \bibinfo {author} {\bibfnamefont{G.}~\bibnamefont{Aubrun}}, \bibinfo {author}
  {\bibfnamefont{S.~J.}\ \bibnamefont{Szarek}},\ and\ \bibinfo {author}
  {\bibfnamefont{D.}~\bibnamefont{Ye}},\ }%
  \bibfield{journal}{%
  \bibinfo {journal} {Comm. Pure Appl. Math.}\ }%
  \textbf{\bibinfo {volume} {67}},\ \bibinfo {pages} {129} (\bibinfo {year}
  {2014})%
  \bibAnnoteFile{NoStop}{Aubrun2014}%
\bibitem{vidal}%
  \BibitemOpen
  \bibfield{author}{%
  \bibinfo {author} {\bibfnamefont{G.}~\bibnamefont{Vidal}}\ and\ \bibinfo
  {author} {\bibfnamefont{R.~F.}\ \bibnamefont{Werner}},\ }%
  \bibfield{journal}{%
  \bibinfo {journal} {Phys. Rev. A}\ }%
  \textbf{\bibinfo {volume} {65}},\ \bibinfo {pages} {032314} (\bibinfo {year}
  {2002})%
  \bibAnnoteFile{NoStop}{vidal}%
\bibitem{peres96}%
  \BibitemOpen
  \bibfield{author}{%
  \bibinfo {author} {\bibfnamefont{A.}~\bibnamefont{Peres}},\ }%
  \bibfield{journal}{%
  \bibinfo {journal} {Phys. Rev. Lett.}\ }%
  \textbf{\bibinfo {volume} {77}},\ \bibinfo {pages} {1413} (\bibinfo {year}
  {1996})%
  \bibAnnoteFile{NoStop}{peres96}%
\bibitem{mhorodeckibound}%
  \BibitemOpen
  \bibfield{author}{%
  \bibinfo {author} {\bibfnamefont{M.}~\bibnamefont{Horodecki}}, \bibinfo
  {author} {\bibfnamefont{P.}~\bibnamefont{Horodecki}},\ and\ \bibinfo {author}
  {\bibfnamefont{R.}~\bibnamefont{Horodecki}},\ }%
  \bibfield{journal}{%
  \bibinfo {journal} {Phys. Rev. Lett.}\ }%
  \textbf{\bibinfo {volume} {80}},\ \bibinfo {pages} {5239} (\bibinfo {year}
  {1998})%
  \bibAnnoteFile{NoStop}{mhorodeckibound}%
\bibitem{Hall1998}%
  \BibitemOpen
  \bibfield{author}{%
  \bibinfo {author} {\bibfnamefont{M.~J.}\ \bibnamefont{Hall}},\ }%
  \bibfield{journal}{%
  \bibinfo {journal} {Phys. Lett. A}\ }%
  \textbf{\bibinfo {volume} {242}},\ \bibinfo {pages} {123} (\bibinfo {year}
  {1998})%
  \bibAnnoteFile{NoStop}{Hall1998}%
\bibitem{Mezzadri2007}%
  \BibitemOpen
  \bibfield{author}{%
  \bibinfo {author} {\bibfnamefont{F.}~\bibnamefont{Mezzadri}},\ }%
  \bibfield{journal}{%
  \bibinfo {journal} {Notices AMS}\ }%
  \textbf{\bibinfo {volume} {54}},\ \bibinfo {pages} {592} (\bibinfo {year}
  {2007})%
  \bibAnnoteFile{NoStop}{Mezzadri2007}%
\bibitem{Marko}%
  \BibitemOpen
  \bibfield{author}{%
  \bibinfo {author} {\bibfnamefont{M.}~\bibnamefont{Znidaric}}, \bibinfo
  {author} {\bibfnamefont{T.}~\bibnamefont{Prosen}}, \bibinfo {author}
  {\bibfnamefont{G.}~\bibnamefont{Benenti}},\ and\ \bibinfo {author}
  {\bibfnamefont{G.}~\bibnamefont{Casati}},\ }%
  \bibfield{journal}{%
  \bibinfo {journal} {J. Phys. A: Math. Theor.}\ }%
  \textbf{\bibinfo {volume} {40}},\ \bibinfo {pages} {13787} (\bibinfo {year}
  {2007})%
  \bibAnnoteFile{NoStop}{Marko}%
\bibitem{Coffman}%
  \BibitemOpen
  \bibfield{author}{%
  \bibinfo {author} {\bibfnamefont{V.}~\bibnamefont{Coffman}}, \bibinfo
  {author} {\bibfnamefont{J.}~\bibnamefont{Kundu}},\ and\ \bibinfo {author}
  {\bibfnamefont{W.~K.}\ \bibnamefont{Wootters}},\ }%
  \bibfield{journal}{%
  \bibinfo {journal} {Phys. Rev. A}\ }%
  \textbf{\bibinfo {volume} {61}},\ \bibinfo {pages} {052306} (\bibinfo {year}
  {2000})%
  \bibAnnoteFile{NoStop}{Coffman}%
\end{thebibliography}%
 
\end{document}